\documentstyle[aps,prb,eqsecnum,epsf]{revtex}
\begin{document}
\draft
\preprint{\today}
\title{The Screening Cloud in the k-Channel Kondo Model: Perturbative and Large-k Results}
\author{Victor Barzykin$^1$ and Ian Affleck$^{1,2}$}
\address{Department of Physics$^1$ and Canadian Institute
for Advanced Research$^2$, \\ 
University of British Columbia, Vancouver, BC, V6T 1Z1, Canada}
\maketitle
\begin{abstract}
We demonstrate the existence of a large Kondo screening cloud in the 
k-channel Kondo model using both renormalization group improved 
perturbation theory and the large-k limit.  We study position (r) dependent 
spin Green's functions in both static and equal time cases.  The 
equal-time Green's function provides a natural definition of the 
screening cloud profile, in which the large scale $\xi_K\equiv v_F/T_K$ appears
($v_F$ is the Fermi velocity; $T_K$ the Kondo temperature).
At large distances it consists of both a slowly varying piece and a piece 
which oscillates at twice the Fermi wave-vector, $2k_F$. 
This function is calculated at all r in the large-k limit.  
Static Green's functions (Knight shift or susceptibility) consist only of a 
term oscillating at $2k_F$, and appear to factorize into a function of r 
times a function of T for $rT/v_F \ll 1$, in agreement with NMR experiments.  
Most of the integrated susceptibility comes from the impurity-impurity part 
with conduction electron contributions suppressed by powers of the bare Kondo 
coupling.  The single-channel and overscreened multi-channel cases are rather 
similar although anomalous power-laws occur in the latter case at large r 
and low T due to irrelevant operator corrections.
\end{abstract}
\pacs{PACS numbers: 
75.20.Hr,
75.30.Mb,
75.40.Cx
}

\section{Introduction}

It is well known that the spin-$1/2$ impurity interacting 
antiferromagnetically with a Fermi liquid
is completely screened at zero temperature\cite{hewson}.
This screening is the essence of the Kondo effect\cite{kondo}.  
The question of the screening length is much more subtle. 
Scaling implies, at least dimensionally, that the 
low energy scale of the model, the Kondo temperature 
$T_K \sim D \exp{(-1/\lambda_0)}$, should be
associated with  an exponentially large length scale, $\xi_K =
v_F/T_K$ (Here $\lambda_0$ is the Kondo coupling times density of
states and $D$ is the band width).  According to Nozi\`eres' Fermi 
liquid picture \cite{nozieres}, one could imagine an electron in a 
region of this size which forms a singlet with the impurity spin. 
Note that this is a more dynamical type of screening than that which 
occurs for charge impurities in a Fermi liquid since it involves a 
linear combination of states where the impurity spin and the 
screening electron spin are in either an up-down or down-up configuration.  
In particular, the finiteness of the  susceptibility at $T\to 0$ 
should not be attributed to a static conduction electron polarization 
cancelling the impurity spin polarization.  Rather it results from the 
tendency of the impurity to form a singlet with the screening electron.

Whether or not this large screening cloud 
really exists  has been a controversial subject in the 
literature, and has recently attracted some theoretical
interest \cite{gan,sorensen,BA,zawadowski,Cox}. 
Boyce and Slichter \cite{slichter} had 
performed direct Knight shift measurements of the spin-spin 
correlator at all temperatures and had concluded that there was no 
evidence of the so-called screening cloud. Their measurements, however,
were limited to very low distances (not more than several lattice spacings), 
and therefore could not probe directly any possible crossover 
at the distance scale $\xi_K$. 

To study the screening cloud, we will consider the behavior of spatial 
spin-spin correlation functions, both zero frequency and equal time. 
There are two distance scales in the Kondo problem at finite
temperature,  $\xi_K$ and the thermal scale $\xi_T = v_F/T$. 
On  general scaling grounds, the spatial correlators 
should depend on the ratio of the distance $r$ to these two scales.  
S\o rensen and one of us \cite{sorensen} have suggested a scaling form 
for the $r$-dependent Knight shift, proportional to the zero frequency 
spin susceptibility, 
which has been justified numerically and perturbatively\cite{sorensen,BA}:
\begin{equation}
\chi(r,T) - \rho/2 = \frac{\cos(2k_Fr)}{8 \pi^2 v_F r^2}\,f(r/\xi_K,r/\xi_T).
\label{sf}
\end{equation}
Here we have subtracted the Pauli contribution $\rho/2$; $\rho$ is
the density of states per spin. The g-factor for the magnetic impurities
is not necessarily equal to that of the conduction electrons. This 
is especially the case for some rare earth ions, which have complex
multiplet structure. If we take into account this possibility, 
scaling properties of the local spin susceptibility become not so simple, 
and we will consider them below. The Knight shift in this case is
a sum of two parts, which scale differently\cite{BA}.

A possible objection to the naive concept of the screening cloud 
is based on  sum rule arguments. The integral of the local spin 
susceptibility Eq.(\ref{sf}) is proportional to
the zero - frequency correlator  $\left< S^z_{el} S^z_{tot} \right>$,
where $S^z_{el}= \int d^3 {\bf r} S^z_{el}({\bf r})$ 
is the spin of the conduction electrons,
$S^z_{tot} = S^z_{el} + S^z_{imp}$.
It can be shown that 
there is no net polarization of the conduction 
electrons\cite{Anderson,schrieffer,lowenstein,lesage}, and  
this correlator should vanish in the scaling limit 
($J \rho \rightarrow 0$ with $T_K$ held fixed). 
At $T=0$ this is simply a consequence of the 
ground state being a singlet. 
As remarked above, this does not necessarily imply the absence of the 
screening cloud in the sense of Nozi\`eres 
but only that the screening is a dynamical process.

In order to see the dynamical cloud of conduction electrons 
let us consider a  snapshot of the system, the
equal-time correlators.  Take  
$K(r,T)=\left<S_{el}^z(r,0)S_{imp}^z(0)\right>$  as an example.  
Note that $\left<S^z_{imp}(0) S^z_{imp}(0) \right> = 1/4$ for a spin-$1/2$ 
impurity, while  $\left<S^z_{el}(0) S^z_{tot}(0) \right> = 0$ as
mentioned above.  (Note that for this 
conserved quantity the equal-time and zero frequency Green's functions 
are proportional to each other.)  Thus the correlator  
$\left<S^z_{el} S^z_{imp}\right> = -1/4$; that is $K(r,T)$ obeys a 
 sum rule. $|K(r,0)|$ is a possible definition of the screening
cloud profile.  
 
The ground state properties of spatial correlators are determined
by the Kondo scale only. In general we expect three different scaling
regimes for $\chi(r,T)$ at a given temperature, with the $r$-boundaries 
defined by the thermal and Kondo length scales. The goal of this paper 
is to determine scaling behavior of the spin correlators in these regimes.

   Exponentially large length scale $\xi_K$, if present, could have
important consequences for the theory of alloys with magnetic impurities. 
Indeed, typical $T_K \sim 10K$ and $E_F \sim 10eV$ makes $\xi_K \sim 10,000a$, 
where a is the lattice spacing. 
Recently this issue was addressed in 1D for Luttinger liquids with 
magnetic impurities\cite{zachar}, where it was found that 
a crossover  happened for $n_{imp} \sim 1/\xi_K$. 

Although perturbative calculations had been done
early on\cite{early}, no definite predictions were 
made regarding the size of the Kondo screening cloud. 
Chen {\em et al.}\cite{chen} have developed renormalization group 
approach. They, however, only considered short-range 
correlations $r \ll \xi_K$.  We use the RG-improved perturbative technique,
 which cannot access lowest temperatures $T<T_K$.
In order to gain some insight into what happens at
low temperatures, we also consider
overscreened $S_{imp}=1/2$ multi-channel Kondo effect, where the 
low-temperature fixed point is accessible perturbatively using 
$1/k$ expansion, k being the multiplicity of the bands. A very 
thorough $1/k$ analysis of the multi-channel Kondo effect has
been performed earlier by Gan \cite{gan}, who, however, came to conclusions 
opposite from ours. We also use the recent conformal field theory approach 
of one of us and Ludwig \cite{affleck,ludwig,AL} to calculate the properties 
of the low-temperature, long distance 
correlation functions and the crossover 
at $\xi_T$. This approach, is valid for all k 
but only for $r \gg \xi_K$, $T \ll T_K$ 
and fails to predict the behavior of the spin correlators 
inside the screening cloud $r \lesssim \xi_K$. 
The result is nevertheless interesting because, as one could 
expect, the spin-spin correlators reflect the non-Fermi-liquid
nature of the overscreened multi-channel fixed point.

The paper is organized as follows. 
In Section II we introduce the model and remind 
the reader how it is transformed to an equivalent 1D model. 
We also define notations which we plan to use in the rest of 
the paper and derive the scaling equations for 
the spin susceptibility. Section III provides detailed 
perturbative analysis of the 
spin-spin correlation functions in the ordinary Kondo model
(The Fermi liquid fixed point).  Section IV is devoted to the  
Non-Fermi-liquid overscreened large-k case, where 
it is possible to obtain results  for the spin correlators at all
temperatures and distances using the $1/k$ expansion. 
We discuss our main conclusions in Section V. 
In Appendix A we mention a few details of our perturbative calculations.  
Appendix B gives the proof of the vanishing of the uniform part 
of the susceptibility.  Appendix C gives results on the overscreened 
case ($k>1$) at $T \ll T_K$ and $r \gg \xi_K$ 
obtained from conformal field theory. 
Some of these results were presented briefly in Ref.[\onlinecite{BA}].
\section{The Model, Renormalization Group and Scaling Equations}

\begin{figure}
\centerline{\epsfxsize=4in\epsfbox{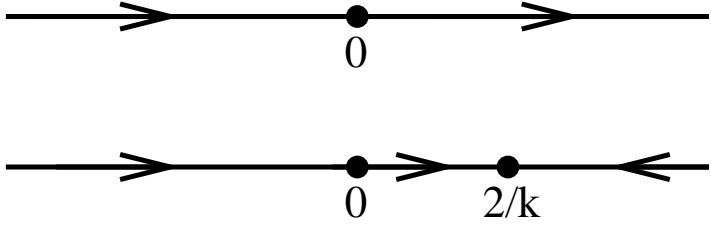}}
\vspace{3mm}
\caption{RG flows for the single-channel and the multi-channel Kondo
problems.}
\label{flow}
\end{figure}

In what follows we consider the standard $S_{\rm imp}=1/2$ Kondo model,
\begin{equation}
H =
\sum_{\bf k}\epsilon_k \psi^{\dagger \alpha}_{\bf  k}\psi_{{\bf k}
\alpha} + J{\bf S}_{\rm imp}\cdot \sum_{\bf  k,\bf  k'}\psi^{\dagger
\alpha}_{\bf  k} \frac{\bbox{\sigma}^{\beta}_{\alpha}}{2}\psi_{{\bf  k'}
\beta}
\label{eq:hkk},
\end{equation}
and the multi-channel $S_{\rm imp}=1/2$ Kondo model.
The Hamiltonian for the $S_{\rm imp}=1/2$ k-channel Kondo model
also includes summation over different channels $j$: 
\begin{equation}
H =
\sum_{{\bf k} j}\epsilon_k 
\psi^{\dagger \alpha j}_{{\bf  k}}\psi_{{\bf k} \alpha j} + 
J{\bf S}_{\rm imp}\cdot \sum_{{\bf  k},{\bf k'}, j}\psi^{\dagger
\alpha j}_{{\bf  k}} 
\frac{\bbox{\sigma}^{\beta}_{\alpha}}{2}\psi_{{\bf  k'}j
\beta}.
\label{lkkm}
\end{equation}
Summation over repeated raised and lowered indiced is implied.
The crucial difference between these two models can be seen from the
form of the $\beta$-function \cite{nozieres}:
\begin{equation}
\beta(\lambda) = - \lambda^2 + {k \lambda^3 \over 2}.
\label{bita}
\end{equation}
The flow of the effective coupling is different (Fig. \ref{flow}) for
$k=1$ and $k>1$. The low temperature fixed point of the multi-channel
Kondo problem is shown to have non-Fermi-liquid nature \cite{affleck}.
At large band multiplicity this nontrivial fixed point becomes
accessible perturbatively.  
This difference is not important for
the purpose of this section, and we use Eq.(\ref{lkkm}) and
Eq.(\ref{bita}) for both multi-channel and $k=1$ models. 

The model is simplified if we assume spherically symmetric Fermi-surface.
Indeed, linearizing the spectrum and observing that scattering only
takes place in the s-wave channel, we can expand the wave functions in
spherical harmonics:
\begin{eqnarray}
\nonumber
\psi_{{\bf k} \alpha j} & = &
\frac{1}{\sqrt{4 \pi} k}\, \psi_{0 \alpha j}(k) +  (...) \\
\nonumber
H_0 & = & \int dk \epsilon(k) \psi^{\dagger \alpha j}_{0}(k)\psi_{0 
\alpha j}(k)  + (...) \\
H_{int} & = & \lambda_0 v_F \int dk dk' \psi^{\dagger \alpha j}_{0}(k)
{ \bbox{\sigma}^{\beta}_{\alpha} \over 2}\, \psi_{0 \beta j}(k')
\cdot{\bf S}_{imp}
\end{eqnarray}  
where $\epsilon(k) = v_F (k-k_F)$ is the linearized spectrum near the Fermi
surface, (...) are higher harmonics. Here $\lambda_0 = \rho J$
is the dimensionless coupling constant of the Kondo model , and
$\rho = k_F^2/ 2\pi^2 v_F$ is the density of states per spin.
 
The s-wave operators obey standard one-dimensional anti-commutation relations,
\begin{equation}
\label{comr}
\left\{ \psi^{\dagger \alpha_1 j_1}_{0} (k), 
\psi_{0 \alpha_2 j_2} (k') \right\} = 
\delta^{\alpha_1}_{\alpha_2} \delta^{j_1}_{j_2}
\delta(k-k').
\end{equation} 
We  define left and right movers on a 
band of width $2 \Lambda$ around $k_F$:
\begin{equation}
\label{lr}
\psi_{L,R} \equiv \int_{- \Lambda}^{\Lambda} d k e^{\pm i k r} \psi_0(k+k_F).
\end{equation}
The 3D fermion operators are then written in the form:
\begin{equation}  
\Psi(r) = \frac{1}{2 \sqrt{2} \pi r} \, \left[
e^{- i k_F r} \psi_L(r) - e^{i k_F r} \psi_R(r) \right] + (...),
\label{3d1d}
\end{equation}
where $(...)$ are higher harmonics.
The left and right-moving fields defined on $r>0$    
obey the boundary condition:
\begin{equation}
\label{bound}
\psi_L(0) = \psi_R(0).
\end{equation}
Flipping the right-moving field to the negative axis, 
$\psi_L(-r) \equiv \psi_R(r)$, we rewrite the 1D Hamiltonian 
in terms of the left-moving field only: 
\begin{equation}
\label{hamil}
H=v_F\int_{-\infty}^\infty
dr\psi^\dagger_L(r)(id/dr)\psi_L(r)
 +  2 \pi v_F\lambda_0
\psi^\dagger_L(0){\bbox{\sigma} \over 2}\psi_L(0)\cdot
{\bf S}_{\rm imp}.
\end{equation}

The purpose of this paper is to analyse various spin-spin correlation
functions. The most important of them is the distance-dependent 
Knight shift, which can be measured in NMR experiments.     
If the impurity spin has a different gyromagnetic ratio from
that of the conduction electrons, the uniform magnetic 
field couples to the spin operator 
${\bf S}_h = {\bf S}_{el} + (g_S/2) {\bf S}_{imp}$,
where ${\bf S}_{\rm imp}$ and ${\bf S}_{el} =
(1/2) \int d{\bf r}\psi^\dagger ({\bf r}) \bbox{\sigma}\psi ({\bf r})$ is
the total spin operator of the impurity and conduction electrons, 
defined with channel sum for the multi-channel problem. 
The expression for the Knight shift then consists of the 
electron and impurity contributions:
\begin{equation}
\chi({\bf r}) \equiv \int_0^{\beta} d \tau < S^z_{el}(r,\tau) S^z_h(0) > = 
\chi_{el}(r)  + {g_s \over 2} \chi_{imp}(r).
\label{ksh}
\end{equation}
(We set $\mu_B=1$.) We will also consider the equal-time spin correlator $K({\bf r})$,
defined by:
\begin{equation}
K({\bf r}) \equiv < S^z_{el}(r,\tau=0) S^z_{imp}(0) >.
\end{equation}
The above 1D formalism allows to simplify this expression for large 
$r k_F \gg 1$. Substituting Eq.(\ref{3d1d}) in Eq.(\ref{ksh}), 
we get:
\begin{equation}
\chi_{A}(r,T)   =  {\chi_{2k_F,A }(r) \over 4\pi^2 r^2v_F} \cos (2k_Fr)
+{\chi_{un,A}(r) \over 8\pi^2 r^2v_F},
\label{scalingK}
\end{equation}
where $A$ corresponds to $imp$ or $el$. For $K({\bf r},T)$ we get a similar
expression:
\begin{equation}
\label{KKK}
K({\bf r},T) = {K_{2k_F}(r) \over 4\pi^2 r^2v_F} \cos (2k_Fr) + {K_{un}(r) \over 8\pi^2 r^2v_F}
\end{equation}
 The total electron spin in 1D is:
\begin{equation}
{\bf S}_{\rm el} = {1\over 2\pi}\int_{-\infty}^\infty
dr\psi^\dagger_L(r){ \bbox{\sigma} \over 2}\psi_L(r).
\label{sel}
\end{equation}
The uniform and $2k_F$ parts 
take the form: 
\begin{eqnarray}
\label{Kchi}
\chi_{un,A}(r,T) & \equiv &
v_F \int_0^{\beta} d\tau <[\psi^\dagger_L(r,\tau){\sigma^z\over 2}\psi_L(r,\tau) 
+\psi^\dagger_L(-r,\tau){\sigma^z\over 2}\psi_L(-r,\tau)]S^z_{\rm A}(0)> 
\\ \nonumber
\chi_{2k_F,A}(r,T) & \equiv &
- v_F \int_0^{\beta} d\tau 
<\psi^\dagger_L(r,\tau){\sigma^z\over 2}\psi_L(-r,\tau) S^z_{\rm A}(0)>.
\end{eqnarray}
Expressions for $K_{un}$ and $K_{imp}$ are analogous to those for 
$\chi_{un,imp}$ and $\chi_{2k_F,imp}$ in Eq.(\ref{Kchi}), although they don't 
involve integration over $\tau$.

If the spins of the impurity and
conduction electron have equal gyromagnetic ratio ($g_s=2$), the operator  
$S^z_h$ is the total spin of conduction electrons and impurity, and
is conserved.  The Knight shift is then given by Eq.(\ref{Kchi}), with
$A=tot$. Since the Kondo interaction is local, only boundary ($r=0$) operators
have non-zero anomalous dimensions. 
Thus the conduction electron spin operator $S_{el}(r)$ 
also has zero anomalous dimension, for $r \neq 0$.  
The local spin susceptibility  
then obeys the following RG equation:
\begin{equation}
\label{rg0}
\left[D{\partial \over \partial D} + \beta(\lambda){\partial
\over \partial \lambda} \right]
\chi(T,\lambda,D, rT/v_F) = 0, \end{equation}
where $D$ is the ultra-violet cut-off (the bandwidth), and
$\beta (\lambda )$ is the $\beta$-function. 
S\o rensen and one of us have recently made a conjection \cite{sorensen}, 
supported by perturbative and numerical results, that  in the scaling limit, 
$rk_F\gg 1$, $T\ll E_F$,
the spin susceptibility has the following form: 
\begin{equation}
\chi\left({rT\over v_F}, {T\over T_K}\right) = 
{\chi_{2k_F}\left({rT\over v_F}, {T\over T_K}\right)\over 4\pi^2 r^2v_F}
\cos (2k_Fr)
\label{scaling}
\end{equation}
where $\chi_{2k_F}$ is a universal functions of two
scaling variables\cite{comment}. This form follows directly 
from Eqs.(\ref{Kchi},\ref{rg0}). In general, one expects that there
could be a non-zero phase in Eq.(\ref{scaling}), and a uniform
term. It is easy to see that the phase is zero due to particle-hole 
symmetry\cite{sorensen}. Indeed, under particle-hole transformation
$\psi_L(r) \rightarrow \sigma^y \psi^{\dagger}_L(r)$, so 
${\bf S}_{tot} \rightarrow {\bf S}_{tot}$, 
$\psi^{\dagger}_L(r) \bbox{\sigma} \psi_L(-r)  \rightarrow
\psi^{\dagger}_L(-r) \bbox{\sigma} \psi_L(r)$. Particle-hole symmetry
of Eq.(\ref{scaling}) then requires that the phase is zero.
This is not so for more realistic Hamiltonians, for which
the particle-hole symmetry is broken. For such Hamiltonians there is an
additional phase $\phi$ in Eq.(\ref{scaling}), but this phase does not
renormalize. 
That is, it is essentially constant in the scaling region ($k_F r \gg 1$). 
The fact that the uniform part
of the spin susceptibility is zero is less trivial. 
For the \underline{static} local spin susceptiblity we have proved\cite{BA}
that all graphs in perturbation theory contain certain integrals that 
vanish. These properties hold for the electron and impurity parts of the
local spin susceptibility Eq.(\ref{Kchi}) separately, for both single-channel
and multi-channel Kondo effects (see Appendix B). The uniform 
part and the phase are zero for the Knight shift in case of 
nontrivial gyromagnetic ratio for the impurity spin ($g_S \neq 1$) as well.

Since we consider the problem perturbatively, it is useful to 
express the scaling function Eq(\ref{scaling}) in terms of some 
effective coupling constants at  an energy scale $E$, $\lambda_E$.
This way we eliminate non-universal $T_K$. The energy scales of
interest are the temperature $T$ and the distance energy scale $v_F/r$.
We will denote corresponding effective couplings as $\lambda_T$ and
$\lambda_r$.
Expressions in terms of effective couplings can be easily converted into
those in terms of $T_K$, and vice versa, provided that the $\beta$-function
is known up to the order needed. Indeed,
\begin{equation}
\label{bedef}
{{d \lambda_E}\over{d \ln{\frac{E}{D}\,}}} \equiv \beta(\lambda_E),
\end{equation}
where $D=v_F/\Lambda$ is the bandwidth.      
Therefore, for the effective coupling at two different energy scales
$E$ and $E'$ we have:
\begin{equation}
\int_{\lambda_{E}}^{\lambda_E'} {{d \lambda}\over{\beta(\lambda)}} = \ln
\frac{E'}{E},
\end{equation}
Since $\lambda_{T_K} \equiv 1$ can well be regarded as one of possible
definitions of $T_K$, we have 
\begin{equation}
\label{deftk}
\int_{\lambda_E}^1 {{d \lambda}\over{\beta(\lambda)}} = \ln \frac{T_K}{E}\, 
\end{equation}
and the arguments of the scaling function in Eq.(\ref{scaling}) can 
be replaced by corresponding effective couplings.

The renormalization group equations for various parts of the local spin 
susceptibility in Eq.(\ref{Kchi}) are less trivial.  
Consider first $\chi_{imp}$. Since the Kondo interaction is at the
origin, the fermion bilinear operator has zero anomalous dimension, 
while the operator $S_{imp}$ receives anomalous dimension,\cite{migdal}
$\gamma_{imp} \simeq \lambda^2/2$.
Renormalizability implies that the functions $\chi_{B,imp}$ ($B = 2k_F$,
$un$) obey equations of the form:
\begin{equation}
\label{rg}
\left[D{\partial \over \partial D} + \beta(\lambda){\partial
\over \partial \lambda} + \gamma_{imp}(\lambda) \right]
\chi_{B,imp}(T,\lambda,D, rT/v_F) = 0, \end{equation} 
where $\gamma_{imp}(\lambda )$ is the anomalous
dimension, which in this case is equal to the 
anomalous dimension of the impurity spin operator.
The other correlator, $\chi_{B,el}$, contains the total conduction
electron spin operator $S_{el}$, in which integration over
the electron spin includes a potentially dangerous region 
near the impurity site. In this region operator mixing occurs 
between the electron spin and impurity spin.  Thus  
$\chi_{B,el}$ obeys non-trivial mixed RG equation:
\begin{equation}
\label{mixed}
\left[D{\partial \over \partial D} + \beta(\lambda){\partial
\over \partial \lambda} \right]
\chi_{B,el}(T,\lambda,D, rT/v_F) = \gamma_{imp}(\lambda) 
\chi_{B,imp}(T,\lambda,D, rT/v_F).
\end{equation}
This equation can be obtained by subtracting Eq.(\ref{rg}) from 
Eq(\ref{rg0}). It is more convenient to express the Knight shift for
$g_S \neq 1$ in terms of $\chi_{imp}$ and $\chi_{tot}$, which obey 
ordinary scaling equations Eqs.(\ref{rg}),(\ref{rg0}). 

The solution of the scaling equation for $\chi_{B,imp}$ Eq.(\ref{rg}) 
has the following form:
\begin{eqnarray}
\label{scE}
\chi_{B,imp}\left(\lambda_0,{T \over T_K},{rT \over v_F}\right) & = &
e^{\int_{\lambda_0}^{\lambda_T}
{\gamma_{imp}(\lambda) \over \beta(\lambda)} d \lambda }
\Pi_{B,imp}\left(\lambda_T,{rT \over v_F}\right) \\ \nonumber
 & = & \Phi_{B,imp}\left(\lambda_T,{rT \over v_F}\right) 
e^{-\int_{0}^{\lambda_0}{\gamma_{imp}(\lambda) 
\over \beta(\lambda)} d \lambda }.
\end{eqnarray}
Here $\Phi_{B,imp}(\lambda_T,rT/v_F)$, $\Pi_{B,imp}(\lambda_T,rT/v_F)$ are 
some scaling functions to be determined below; 
$\lambda_0=\rho J$ is the bare coupling constant. 
The solution of the scaling equations for
$\chi_{imp}$  is a function of $T/T_K$ and $rT/v_F$, up to 
some non-universal coefficient. We see that the non-universal 
coefficient $exp[-\int_{0}^{\lambda_0} d \lambda(\gamma_{imp}(\lambda) / 
\beta(\lambda))]$ is equal to unity in the scaling limit of zero bare
coupling $\lambda_0 \rightarrow 0$, 
if $\gamma_{imp}(\lambda_0)/\beta(\lambda_0)$
is non-singular in this limit. This is indeed the case for the Kondo model.
The scaling function $\Phi_{B,imp}(T/T_K,rT/v_F)$, of course, can differ from
$\chi_{2k_F}(T/T_K,rT/v_F)$ in Eq.(\ref{scaling}).  The equal-time
correlation functions also obey analogous scaling equations Eq.(\ref{rg})
with the anomalous dimension which is a sum of the dimension of the 
corresponding operators (for $\chi_{imp}$ it is again $\gamma_{imp}$) . 
These equations are also applicable to the uniform part of the correlator, 
which is now non-zero.  

  In the rest of this paper we will consider these scaling functions
in various regimes, which we now outline. 
 Scaling form is applicable for $r \gg 1/k_F$, and $T \ll D \approx E_F$.  
For the single-channel Kondo model perturbative treatment is  
only valid for $T \gg T_K$. 
From Eq.(\ref{sf}) one could expect 
that there could be two  
crossovers: one at $\xi_T$ and one at $\xi_K \gg \xi_T = v_F/T$.
The latter crossover, however, 
does {\em not} happen as a function of $r$ for $T \gg T_K$.  
The low-temperature correlation functions in
the single-channel Kondo effect can be studied  using the Fermi liquid 
approach\cite{nozieres}. The region of validity for this approach  is 
$r \gg \xi_K$, $T \ll T_K$. It provides important information about 
the low-temperature long-distance form of the correlation functions, 
and the crossover at $\xi_T \gg \xi_K$, but is unable to access 
the most interesting region $r \sim \xi_K$, and answer the question 
of existence of the screening cloud. For the multi-channel 
Kondo effect the low-temperature long 
distance correlation functions can be obtained
using the conformal field theory approach \cite{affleck,ludwig,AL}, which is a
generalizarion of Nozi\`eres' Fermi liquid picture. 
It is also limited to $r \gg
\xi_K$, $T \ll T_K$.   The interesting low temperature region 
with $r \sim \xi_K$ only becomes accessible at large k, when 
the whole scaling function can be constructed. 

\section{The single-channel Kondo model.}
\subsection{The Knight shift.}
In what follows we consider the Knight shift in the single channel
Kondo model. As mentioned above, the local spin susceptibility 
only has the oscillating part. We have calculated it up to     
the third order in perturbation theory\cite{BA}. Summing all the relevant
diagrams (see Appendix A), we obtain: 
\begin{eqnarray} 
\label{Pth}
\chi_{2k_F}\left(x={rT\over v_F},\lambda_0,D\right) & = &
\frac{\pi^2}{4 \sinh(2 \pi x)} \, [\lambda_0 + \lambda_0^2
(\ln(D/T) + M(x)+x) \\ \nonumber
& + & \lambda_0^3 (\ln^2(D/T)+ \ln(D/T)(2 M(x)+ 2 x - 0.5) \\ \nonumber
& + & (M(x)+x) (M(x) + 0.5) + const) ],
\end{eqnarray}  
where 
\begin{equation}
\label{mofx}
M(x)=\ln\left[1-\exp(-4 \pi x) \right].
\end{equation}
Substituting this expression in Eq.(\ref{rg}) we find that scaling is
indeed obeyed. At small $r$, $x \ll 1$, Eq.(\ref{Pth}) is rewritten as:
\begin{eqnarray}
\label{lrdiv}
\chi_{2k_F}(r,\lambda_0,D) & = & {\pi v_F \over 8 r T} [\lambda_0
+\lambda_0^2 \ln(\tilde{\Lambda} r) + \lambda_0^3 \ln^2(\tilde{\Lambda} r) + 
0.5 \lambda_0^3 \ln(\tilde{\Lambda} r) \\ \nonumber
& - & \lambda_0^3 \ln(D/T) + \lambda_0^3 const], 
\end{eqnarray}
where $\tilde{\Lambda} = 4 \pi D/v_F = 4 \pi \Lambda \sim k_F$.
It is clear from Eq.(\ref{lrdiv}) that 
the infrared divergences of the perturbation theory {\it are not} cut
off at low T by going to small $r$, as was first noticed by
Gan \cite{gan}. In the third order, these
divergences are associated with the graph shown in Fig.2.  Due to the
non-conservation of momentum by the Kondo interaction, the
bubble on the right gives a logarithmic $T$-dependent factor which is
independent of $r$.
Thus, the interior of the screening cloud does {\it not} exhibit
weak coupling behavior.

It is convenient to rewrite this
result in terms of effective couplings at the energy scales $T$ and $v_F/r$.  
One can easily write down the effective coupling constant\cite{abrikosov} 
at some energy scale $\omega$ using the well-known 
$\beta$-function Eq.(\ref{bita}):  
\begin{equation} 
\label{effc}\lambda_{\omega}=\lambda_0 +
\lambda_0^2\ln (D/\omega)+
\lambda_0^3[\ln^2(D/\omega)-(1/2)\ln (D/\omega)+constant].
\label{lambdaeff}\end{equation}
We find that the expression for $\chi_{2k_F}$ 
is simplified when we use effective
couplings $\lambda_T$ and $\lambda_E$ at the 
energy scales $T$ and $E(x) = T/[1-\exp(- 4 \pi x)]e $,
$x = r/\xi_T$. When $r \ll \xi_T$ the latter 
becomes the effective coupling at 
the distance scale $r$, since $E(x) \propto v_F/r$.  
Eq.(\ref{Pth}) in terms of these effective couplings takes the form:
\begin{equation}
\chi_{2k_F}\left(x={rT\over v_F},\lambda_T\right) = 
{(\lambda_E+(3\pi/2) \lambda_E^2 x + const \lambda_E^3)(1-\lambda_T) 
\over (4/\pi^2 )\sinh(2\pi x)},
\label{p3}
\end{equation}

It is instructive to consider various limiting cases for the scaling function 
Eq.(\ref{p3}). For $r \ll \xi_T$  we find:
\begin{equation}
\chi_{2k_F}(x,\lambda_T) = 
(\pi/8x)(\lambda_r +const \lambda_r^3)(1-\lambda_T).
\label{c}
\end{equation}
If  $r \gg \xi_T$, the spin susceptibility takes the following form:
\begin{equation}
\chi_{2k_F}(x,\lambda_T) = (3 \pi^3/4) \lambda_T^2(1-\lambda_T) e^{-2 \pi x}.
\end{equation}
For high temperatures $T \gg T_K$ there is no crossover  
at $r \sim \xi_K$ in the behavior of the local spin susceptibility. 
The factor $(1-\lambda_T)/4T$ in Eq.(\ref{c}) is,
to the order under consideration, precisely the total impurity susceptibility,
$\chi_{\rm tt}(T)$.  This is the total susceptibility less the bulk Pauli
term and its value has been determined accurately \cite{hewson}. Thus 
Eq.(\ref{c}) can be written:
\begin{equation}
\chi_{2k_F}(T,r) = {(\lambda_r+const\lambda_r^3)
\over  2 (r/\pi v_F)} \chi_{\rm tt}(T),  
\label{fac}
\end{equation}
We can compare this result with the experiment of
Boyce and Slichter\cite{slichter}, who have measured the Knight
shift from Cu nuclei near the doped Fe impurities, at distances
up to 5-th nearest neighbor. 
At these very small distances of order of a few lattice spacings,
they have found empirically that the Knight shift obeyed a
factorized form, $\chi(r,T) \approx f(r)/(T+T_K)$,
with rapidly oscillating function $f(r)$ for a
wide range of $T$ extending from well above to well below the Kondo
temperature. Although our condition $r \gg 1/k_F$ is not satisfied
in this experiment, this form coincides with Eq. (\ref{fac}),
since the Bethe Ansatz solution for $\chi_{tt}(T)$ may be quite 
well approximated\cite{hewson} by $1/(T+T_K)$ at intermediate temperatures
$T \sim T_K$. As one can see from Eq.(\ref{p3}), 
this factorization breaks down  at $r \sim v_F/T$. 

 At low temperatures $T \ll T_K$ and large distances $r \gg \xi_K$
the behavior of $\chi(r)$ is determined by the zero-energy Fermi
liquid fixed point\cite{sorensen}. The Kondo impurity acts as 
a potential scatterer with a phase shift $\pi/2$ at the Fermi 
surface\cite{nozieres}. The local susceptibility follows directly
from the formula for Friedel's oscillation in the electron density
for an s-wave scatterer and $\pi/2$ phase shift,
\begin{equation}
n(r) = n_0 - \frac{1}{2 \pi^2 r^3}\, cos[2 k_F r + \pi/2].
\end{equation} 
Since the magnetic field $H$ simply shifts the chemical potential
by $\pm g \mu_B H/2$ for spin up or spin down electrons, 
\begin{equation}
\chi(r,T) = {1 \over v_F} {dn \over {dk_F}} = {\rho \over 2} + 
{1 \over {4 \pi^2 v_F r^2}} cos(2 k_F r).
\end{equation}
This implies for the scaling function Eq.(\ref{rg}):
\begin{equation}
\label{trivial}
\chi_{2k_F} = 1.
\end{equation}

The finite-temperature properties of $\chi_{2k_F}(r)$, and, in
particular, the crossover at $r \sim \xi_T$ can be obtained
directly from the Nozi\`eres' low-energy Hamiltonian for the 
Fermi liquid fixed point\cite{nozieres,affleck}:
\begin{equation}
H_0 = \int_{- \infty}^{+ \infty} d r \psi_L^{\dagger}(r)
{d \over {dr}} \psi_L(r) + {\delta(r) \over T_K} {\bf S}^2_{el}(r),
\end{equation}
where ${\bf S}_{el}(r) \equiv \psi^{\dagger}_L(r) (\bbox{\sigma}/2)
\psi_L(r)$. This definition of $T_K$ differs from one in
Eq.(\ref{deftk}) or $\chi_{tt} \propto 1/(T+T_K)$ by numerical factors $O(1)$
(Wilson ratio). The expression for $\chi_{2 k_F}(r)$, 
Eq.(\ref{trivial}), is zero order 
in the leading irrelevant coupling constant $1/T_K$, and the
finite-temperature form of $\chi_{2k_F}(r)$ is easily obtained:
\begin{equation}
\chi_{2k_F}(x) = {2 \pi x \over {\sinh(2 \pi x)}},  \ \ x = {r T \over v_F}.
\label{chlow}
\end{equation}
We can derive corrections to Eq.(\ref{chlow}) by doing perturbation
theory in the leading irrelevant operator. For the first correction
we obtain:
\begin{equation}
\delta \chi_{2k_F}(x) = {\pi^2 T \over {T_K \sinh(2 \pi x)}}.
\end{equation}
The first correction does not alter the leading order behavior.
At zero temperature the scaling function for $r \gg \xi_K$ takes 
the following form:
\begin{equation}
\label{FLR}
\chi_{2k_F}(r/\xi_K) = 1 + \pi {\xi_K \over 2 r}.
\end{equation}
This correction gives rise to the first
term in the large-distance expansion of our scaling function
$\chi_{2k_F}(r/\xi_K)$.  
The behavior of the scaling function $\chi_{2k_F}$ in 
different regimes is summarized in Fig.(\ref{diagr1}). 
$\chi_{2k_F}(r/\xi_K,r/\xi_T)$
exhibits a crossover at low $T$, when the 
``screening'' cloud is formed. At high temperatures this
crossover is absent. 

What happens when the $g$-factor of the impurity is anomalous?
$\chi_{2k_F}(r,T)$ is a sum of impurity and electron parts,
$\chi_{2k_F,\rm{imp}}$ and $\chi_{2k_F,\rm{el}}$. As we have
discussed in the previous section, the latter obeys a complicated 
mixed RG equation, Eq.(\ref{mixed}). It is more convenient
to express the spin susceptibility in terms of the correlators
$\chi_{2k_F,{\rm tot}}(r,T)$ and $\chi_{2k_F,{\rm imp}}(r,T)$,
for which RG equations are simple:
\begin{eqnarray}
\chi_{2k_F}(r,T) &=& 
(g_S/2) \chi_{2k_F,\rm{imp}}(r,T) + \chi_{2k_F,\rm{el}}(r,T) \\
\chi_{2k_F,\rm{el}}&=& \chi_{2k_F,\rm{tot}}(r,T) - \chi_{2k_F,\rm{imp}}(r,T)
\nonumber
\end{eqnarray}
Since we have already determined $\chi_{2k_F,\rm{tot}}(r,T)$, it is
sufficient to consider only $\chi_{2k_F,\rm{imp}}(r,T)$.
From the perturbative analysis (see Appendix A) we obtain:
\begin{eqnarray}
\label{Pthimp}
\chi_{2k_F,{\rm imp}}\left(x={rT\over v_F},\lambda_0,D\right) & = &
\frac{\pi^2}{4 \sinh(2 \pi x)} \, [\lambda_0 + \lambda_0^2
(\ln(D/T) + M(x)+0.5) + \\ \nonumber
 & &\lambda_0^3 (\ln^2(D/T) + 2 \ln(D/T) M(x) + (M(x) + 0.5)^2 + const) ],
\end{eqnarray}
where $M(x)$ is the same as in Eq.(\ref{mofx}). One can easily check that
Eq.(\ref{rg}) is obeyed with\cite{abrikosov,gan}:
\begin{equation}
\beta(\lambda) = - \lambda^2 + {\lambda^3 \over 2}, \ \
\gamma_{\rm imp}(\lambda) = {\lambda^2 \over 2}.
\label{bl}
\end{equation}
We then obtain for the non-universal factor in Eq.(\ref{scE}):
\begin{equation}
\label{fct}
e^{-\int_{0}^{\lambda_0}{\gamma_{imp}(\lambda) 
\over \beta(\lambda)} d \lambda } \simeq 1 + {\lambda_0 \over 2},
\end{equation} 
and the local impurity spin susceptibility takes the following
form:
\begin{equation}
\chi_{2k_F,\rm{imp}}  
\simeq \left(1 + {\lambda_0 \over 2} \right) \chi_{2k_F}^{(1)}(\lambda_T,x),
\end{equation} 
where the scaling function
\begin{equation}
\chi_{2k_F}^{(1)}(\lambda_T,x)=
{(\lambda_E+const \lambda_E^3)(1-\lambda_T)\over
(4/\pi^2 )\sinh(2\pi x)}
\end{equation}
differs from that for the conserved local susceptibility Eq.(\ref{p3}). 
For the electron part we obtain:
\begin{equation} \chi_{2k_F,\rm{el}} \approx 
-{\lambda_0 \over2} \chi_{2k_F}^{(1)}(\lambda_T,x)
+ {(3\pi/2) \lambda_E^2 x(1-\lambda_T) \over
(4/\pi^2 )\sinh(2\pi x)}.
\label{con2k_F}\end{equation}
The second contribution does not vanish in the scaling limit 
$\lambda_0 \rightarrow 0$. However, it only becomes substantial at 
large distances $r \sim \xi_T$, where there is no additional
smallness associated with the factor $x = r/\xi_T$ . 
We conclude that two different scaling functions are present in the
experimentally measured Knight shift, and their share depends upon
the gyromagnetic ratios for the impurity and the conduction
electrons. 

\subsection{Integrated susceptibilities}

It is instructive to consider the integral of 
$\chi({\bf r},T)$ over all space.
This quantity determines the polarization of the 
screening cloud in external magnetic field. We immediately see that
the contribution from large distances vanishes because of the oscillatory
behavior of $\chi({\bf r},T)$ at large $r$. Nevertheless the integral
can be finite  due to the contributions at small distances $r \sim 1/k_F$.
We will specify three different spin correlators: 
\begin{eqnarray}
\label{susc}
\chi_{\rm tt}(T) &\equiv& {<S^z_{\rm tot} S^z_{\rm tot}>\over T}, \ \ 
\chi_{\rm ti}(T) \equiv 
{<S^z_{\rm imp} S^z_{\rm tot}>\over T},\nonumber \\   
\chi_{\rm ii}(T) &\equiv& \int_0^\beta 
<S^z_{\rm imp}(\tau ) S^z_{\rm imp}(0)>  
d\tau .\end{eqnarray}
For this choice of correlators the RG equations are simplified, and have 
the form Eq.(\ref{rg}).
It seems more natural to define correlators of the impurity
spin and the total conduction electron spin $S_{el}$ instead of $S_{tot}$:
\begin{eqnarray}
\chi_{\rm ee}(T) &\equiv& \int_0^\beta <S^z_{\rm el}(\tau ) S^z_{\rm
el}(0)> d\tau - \chi_0\\
\chi_{\rm ei}(T) &\equiv& \int_0^\beta <S^z_{\rm el}(\tau ) S^z_{\rm
imp}(0)> d\tau , \nonumber
\end{eqnarray}
where $\chi_0$ is the free electron susceptibility, proportional to
the volume of the system.
However, for this set of spin correlators the RG equations are mixed.

Two of the three spin correlation functions can be measured. The first
one is the bulk susceptibility, $\chi_{tt}(T)$.
The electron spin polarization in the presence
of an impurity  is determined by the spatial integral of $\chi(r)$ measured
in the Knight shift experiment Eq.(\ref{ksh}),
or, equivalently, by $\chi_{\rm tt}(T)-\chi_{\rm ti}(T)$. 
If the gyromagnetic ratio for the impurity is different from 2,
the experimentally measured magnetic
susceptibility is $(g_s^2/4) \chi_{\rm ii} + g_s \chi_{ie} + \chi_{ee}$, 
while the integrated electron susceptibility is given
by $(g_s/2) \chi_{ie}+\chi_{ee}$.

Since $S^z_{\rm tot}$ is conserved, the spin susceptibilities obey
the RG equation, Eq. (\ref{rg}), with anomalous dimensions determined
by the dimension $\gamma_{\rm imp}(\lambda)$ of the operator
$S^z_{\rm imp}$.  For the three different susceptibilities:
$\gamma_{\rm tt}=0$,
$\gamma_{\rm ti}=\gamma_{\rm imp}$, 
and $\gamma_{\rm ii} = 2 \gamma_{\rm imp}$. 
The solutions of these equations take the form Eq.(\ref{scE}),
\begin{equation}
\label{solu}
4 T \chi_j(T) = \Phi_j(\lambda_T) e^{-\int_{0}^{\lambda}{\gamma_j(\lambda)
\over \beta(\lambda)} d \lambda },
\end{equation}
where $j$ labels ${\rm tt}$, ${\rm ti}$ or ${\rm ii}$. 
From our third-order perturbative analysis using Wilson's 
result\cite{wilson} for $\chi_{tt}(T)$ we have obtained that 
the functions $\Phi_j(\lambda_T)\simeq 1-\lambda_T$ 
coincide for all three susceptibilities up
to and including  terms of order $\lambda_T^2$.  If this is indeed the case
in the Kondo model, we then obtain from Eq.(\ref{solu}) that in the scaling
limit $\lambda_0 \rightarrow 0$ both $\chi_{ee}(T)$ and $\chi_{ie}(T)$ vanish.
At finite bare couplings these susceptibilities also become finite, with
non-universal amplitudes. We then obtain from Eq.(\ref{solu}) for the 
impurity-electron and electron-electron pieces of the spin susceptibility: 
\begin{eqnarray}
\chi_{ie} &\simeq& - {\lambda_0 \over 2}  \chi_{\rm{tt}}(T) \\ \nonumber
\chi_{ee} &\simeq& {\lambda_0^2 \over 4}  \chi_{\rm{tt}}(T),
\end{eqnarray}
Thus, the integrated distance-dependent Knight shift obeys:
\begin{equation}
\int \chi(r,T) d{\bf r} = \chi_{ee}(T)+\chi_{ie}(T) 
\approx - {\lambda_0 \over 2}  \chi_{\rm{tt}}(T).
\label{conint}\end{equation}  
The major contribution to Eq.(\ref{conint}) comes from the
electron-impurity correlator.
It should be emphasized
that the result is non-zero at finite bare coupling $\lambda_0$. (A typical
experimental value of $\lambda_0$ might be $1/\ln (E_F/T_K)\approx .15$.)  
It is easy to see that the integral in Eq.(\ref{conint}) is dominated
by $r \sim 1/k_F$. Thus most of the small net polarization of the
electrons in a magnetic field (with the free electron value subtracted)
comes from very short distances. However, this should not be interpreted 
as meaning that the screening cloud is small as can be clearly seen from
the equal-time correlation function discussed in the next sub-section.

If the equality of the scaling functions
$\Phi_j(\lambda_T)$ defined in Eq. (\ref{solu})
holds at all T, the integrated electron spin
susceptibility vanishes in the scaling limit of zero bare coupling at
all $T$. The fact that $\chi_{ie}$ and $\chi_{ee}$ are suppressed in the 
scaling limit has been known or conjectured from a variety of 
different approaches over the years.  The earliest result of 
this sort that we are aware of, in the context of the Anderson 
model, predates the discovery of the Kondo effect and is 
referred to as the Anderson-Clogston compensation theorem\cite{Anderson}.
It was later established at T=0 from the Bethe ansatz 
solution\cite{lowenstein}.  A very simple and general proof\cite{lesage} 
of this result follows from the abelian bosonization approach\cite{emery}.  
Beginning with left-moving relativistic fermions on the entire real line, 
as in Eq. (\ref{hamil}), we may bosonize to obtain left-moving 
spin and charge bosons.  The charge boson decouples and  the Hamiltonian 
for the Kondo Hamiltonian can be written in terms of the left-moving 
spin boson, $\phi_L$ which obeys the canonical commutation relation:
\begin{equation}
[\phi_L(r'),\partial_r\phi_L(r)]=(i/2)\delta (r-r').
\end{equation}
The Hamiltonian becomes:
\begin{eqnarray}
H &=& \int_{-\infty}^{\infty} dr \left[v_F (\partial_r \varphi_L(r) )^2 - {h_e \over 
\sqrt{2 \pi}} \partial_r \varphi_L(r) \right] + H_K - h_i S^z \\ \nonumber
H_K &=& 2 \pi v_F\lambda_0 \left[{S^z \partial_r \varphi_L(0) \over \sqrt{2 \pi}} +
const \cdot ( S^+ e^{i \sqrt{8 \pi} \varphi_L(0)} + h.c.) \right]
\end{eqnarray}
Here $h_i$ and $h_e$ are the magnetic fields acting on the
impurity and the conduction electrons, correspondingly. These fields may
differ by the ratio of corresponding g-factors. 
We can get rid of the $\int dr  \partial_r \varphi_L(r)$ term by shifting
the bosonic field:
\begin{equation}
\varphi_L(r) = \tilde{\varphi}_L(r) + h_e r /v_F\sqrt{8 \pi}
\end{equation}
The Hamiltonian in terms of the new bosons takes the form:
\begin{eqnarray}
H &=& \int_{-\infty}^{\infty} dr 
v_F(\partial_r \tilde{\varphi}_L(r) )^2 - {h_e^2 L \over
4 \pi v_F} + \sqrt{2 \pi} v_F\lambda_0 S^z \partial_r \tilde{\varphi}_L(0) \\ \nonumber
& &+ const \cdot 2 \pi \lambda_0  ( S^+ e^{i \sqrt{8 \pi} \tilde{\varphi}_L(0)} 
+ h.c.) - \left(h_i- {h_e \lambda_0 \over 2}\right) S^z.
\end{eqnarray}

Thus, our original Hamiltonian with non-zero field $h_e$ acting on the
conduction  electrons is exactly equivalent to the one with no field acting
on conduction electrons and modified 
impurity field. The same argument was given in Ref.[\onlinecite{Lesage}] 
except that the field shift by $h_e\lambda_0/2$ was not obtained because 
another, non-commuting, canoncial transformation was performed first 
to eliminate the $z$-component of the Kondo interaction.

In terms of the free energy, this is written as:
\begin{equation}
F(h_i,h_e) = - {h_e^2L\over 4 \pi v_F}   + F(0, h_i- {h_e \lambda_0 \over 2})
\end{equation}

Taking  magnetic field derivatives, we easily find:
\begin{eqnarray}
\chi_{ii} &=& - {\partial^2 F \over \partial h_i^2} \\ \nonumber
\chi_{ie } &=& - {\partial^2 F \over \partial h_i \partial h_e} 
= - {\lambda_0 \over 2} \chi_{ii} \\ \nonumber
\chi_{ee} &=& - {\partial^2 F \over \partial h_e^2}-\chi_0 = 
\left({\lambda_0 \over 2}\right)^2 \chi_{ii}, 
\end{eqnarray}
where $\chi_0 = {L \over 2 \pi v_F}$ is the Pauli term.
It is easy to see that this is valid for the anisotropic Kondo model as well, 
with $\lambda_0$ being the z-component of the Kondo interaction.

\subsection{Equal-time spin-spin correlator}

      The equal-time spin correlators provide a snapshot
of the Kondo system. The quantity of interest is: 
\begin{equation}
K({\bf r},T) = \left< S^z_{el}({\bf r},0) S^z_{imp}(0) \right> .     
\label{etc}
\end{equation}
As we have shown in the previous sections, it satisfies a non-zero sum rule:
\begin{equation}
\int d{\bf r} K({\bf r},T) = -1/4.
\label{smr}
\end{equation}
The proof that $\chi_{un}(r) = 0$ is based on the fact that the time integral
for the Feynman diagrams is zero in all orders in perturbation theory 
(see Appendix B). For the equal-time correlator we don't integrate over 
the time variable, so the uniform part does not have to vanish. $K({\bf r},T)$
can be rewritten in 1D in terms of the uniform and $2k_F$ parts,
Eqs(\ref{scaling},\ref{Kchi}). 
For the same reason as for the impurity part of the
Knight shift, the equal-time correlator obeys the scaling equation 
Eq.(\ref{rg}), with solutions of the form Eq.(\ref{scE}). Since the
decomposition of $K(r,T)$ into the uniform and $2k_F$ parts is only valid
in the scaling region $k_F r \gg 1$, the sum rule Eq.(\ref{smr}) does
not necessarily extend to $K_{un}(r,T)$. The region $r \sim 1/k_F$ could
produce a large contribution  to the sum rule Eq.(\ref{smr}).

 Consider now the equal-time correlators $K_{un}(r,T)$ and $K_{2k_F}(r,T)$
perturbatively. In the third order we obtain (see Appendix A):
\begin{eqnarray}
\label{etp}
K_{un}(r,T) &=& 
{\pi^2 (-\lambda_0^2+(1/2)\lambda_0^3 - 2 \lambda_0^3 \ln [D/T] ) T 
\over {\exp(2 \pi r/\xi_T) - 1}} +  \pi^2 T \lambda_0^3 G_1(r/\xi_T)\\ 
\nonumber
K_{2k_F}(r,T)- T \chi_{2k_F}(r,T) &=& \pi^2 \lambda_0^3 T  
 \left[G_2(r/\xi_T)-{\ln(1-e^{-2\pi r/\xi_T}) \over 
{4\sinh(2 \pi r/\xi_T)}}\right].
\end{eqnarray}
$G_{1,2}(x)$ are some functions which can be represented 
as integrals:
\begin{eqnarray}
\label{g1g2}
G_1(x) &=& \int_0^1 {2 ds \over {1-s}}
\left[ {1 \over {e^{2 \pi x} -1 }} + {s \over {1-s} }
\ln \left( {1-e^{-2 \pi x} \over {1 - s e^{-2 \pi x}}} 
\right) \right] \\ \nonumber
G_2(x) &=& \int_0^1 ds  
{e^{-2 \pi x} s \over {\left(1-s\right)\left(1-s e^{-4 \pi x}\right)}}
\ln \left[{1-s e^{-2 \pi x} \over {1 - e^{-2 \pi x }}} \right]. 
\end{eqnarray}
It is easy to check that Eq.(\ref{rg}) is satisfied for both uniform 
and $2k_F$ parts. The solutions are found in the form 
Eq.(\ref{scaling}) with the non-universal factor Eq.(\ref{fct}). 
The scaling functions are easily obtained
from Eq.(\ref{etp}). The final expressions are simplified 
in the most interesting limiting cases. For $r \ll \xi_T$ we obtain:
\begin{eqnarray}
\label{keqt}
K_{un} (\lambda_r, r/\xi_T,\lambda_0) &=&   
- {\pi v_F \lambda_r^2 (1 + \lambda_0/2)\over 2r} \\ \nonumber
K_{2k_F}(\lambda_r, r/\xi_T,\lambda_0)&=&
{\pi v_F \lambda_r (1 + \lambda_0/2)} \over {8r}
\end{eqnarray}
In case of $r \gg \xi_T$, these functions take the form:
\begin{eqnarray}
K_{un} (\lambda_r, r/\xi_T,\lambda_0) &=& -\pi^2 T \lambda_T^2 
\left(1 + {\lambda_0 \over 2}\right) e^{-2 \pi r/\xi_T}   \\ \nonumber
K_{2k_F}(\lambda_r, r/\xi_T,\lambda_0) &=& 
{\pi^2 T \lambda_0 \over 2}\left(1 + 
{\lambda_0 \over 2}\right) (1-\lambda_T) e^{-2 \pi r/\xi_T}.
\end{eqnarray}
Note that $K_{2k_F}$ is suppressed in this limit 
by the small value of the bare coupling.
Like the local spin susceptibility, 
the equal-time correlator does not have crossover 
at $r \sim \xi_K$ at high temperatures. 
Instead, the corresponding scaling function
for $r \gg \xi_T$ has a factorized form, 
$K(r/\xi_T,T/T_K) \propto f_1(r/\xi_T) f_2(T/T_K)$. 

 The behavior of the equal-time correlation function at 
$T \ll T_K$ and $r \gg \xi_K$ can be calculated using Nozi\'eres Fermi
liquid approach. Indeed, the impurity spin at the infrared 
Kondo fixed point should be replaced by the local spin density 
${\bf J}(0)$ for $r\approx 0$, up to a constant multiplicative
factor \cite{affleck,ludwig}:
\begin{equation}
{\bf S}_{imp} \propto {v_F {\bf J}_L(0) \over T_K}.
\end{equation}
Substituting this in the definition of $K_{un}$ and $K_{2k_F}$,
we obtain at finite T:
\begin{equation}
K_{2k_F}(r/\xi_T) = - (1/2) K_{un}(r/\xi_T) =  
{const T^2\over T_K \sinh^2(\pi T r/v_F)} 
\end{equation}
Thus, at $T \rightarrow 0$  the equal-time correlator K decays as $\sin^2k_Fr/r^4$
(see Eq.(\ref{KKK})). This result was obtained by Ishii\cite{ishii} in the
context of the Anderson model.
The behavior of the equal-time correlators $K_{2k_F}(r)$ and $K_{un}(r)$
in different regimes is summarized in Fig.\ref{diagr2} and Fig.\ref{diagr3}.

\section{Large k multi-channel Kondo model.}

The information that one gets for the single-channel Kondo model using
perturbative RG is very limited, and further numerical analysis is required.
To justify the presence of the Kondo length scale more, we
analyse the multi-channel model with large band  multiplicity.
The generalization of the above perturbative analysis to the
multi-channel case is quite straightforward. 
The Hamiltonian for the $S_{\rm imp}=1/2$ k-channel Kondo model
is given by Eq.(\ref{lkkm}).
Further analysis of Section II applies 
to the multi-channel case as well.
Some of the relevant perturbative $1/k$ calculations 
for the multi-channel Kondo effect were done by Gan\cite{gan}. 
His scaling equations and conclusions about the screening cloud are, 
however, different from ours.  We refer to some of his results below.   

\subsection{The local spin susceptibility.}

Spin susceptibilities of the multi-channel Kondo problem also satisfy 
RG equation Eq.(\ref{rg}). However, the diagrams which contribute
to the same order in $1/k$ are different from the single-channel case.
Since the low-temperature fixed point for coupling constant 
is $\sim 1/k$,  each vertex produces a $1/k$ factor. [Here we assume that the
bare coupling, $\lambda_0$, is also O(1/k).]
Each loop, on the other hand, gives a large factor of $k$. Combination
of these factors determines the diagrams that one needs to calculate
to a given order in $1/k$. The number of diagrams 
is finite (see Appendix A for details). We shall calculate the spin
correlators of interest up to the first non-zero order in $1/k$.

The solution of the scaling equations for the coupling constant up 
to subleading order in $1/k$ were obtained by Gan \cite{gan}. From
the calculation of the conduction electron self-energy he finds
that the $\beta$-function is given by:
\begin{equation}
\label{beta2}
\beta(\lambda_E) =  - \lambda_E^2 + {1 \over 2} k \lambda_E^3 + {1 \over 2}
k a \lambda_E^4 - {1 \over 4} k^2 \lambda_E^5,
\end{equation}
where a is some non-universal number, which depends on the cutoff procedure.
The flow for the overscreened Kondo model is shown in Fig.\ref{flow}. 
The low-temperature physics is determined by
the intermediate-coupling stable fixed point $\lambda^*$ given 
by $\beta(\lambda^*) = 0$:
\begin{equation}
\lambda^* = {2 \over k} \left(1 + {{2 - 4a} \over k}\right).
\end{equation}
The position of the fixed point is not universal.
On the other hand, the slope of the beta-function at this fixed point, 
$\Delta \equiv \beta'(\lambda^*)$ is the dimension of the leading irrelevant
operator \cite{affleck}, and should be universal:
\begin{equation}
\label{delt}
\Delta = \beta'(\lambda^*) = {2 \over {k+2}}.
\end{equation}
This fact is readily checked from Eq.(\ref{beta2}). 

It is sufficient for our purposes to consider the $\beta$-function in the
leading order in $1/k$, Eq.(\ref{bita}). At this order 
$\lambda^* = \Delta = 2/k$. 
Solving Eq.(\ref{bedef}), we obtain:
\begin{equation}
{k \over 2} (\phi_E + \ln|\phi_E|) = \ln{E \over T_K}, 
\end{equation}
where 
\begin{eqnarray}
T_K &=& D \exp\left[{k \over 2} - {1 \over \lambda_0} \right] 
\left( {k \lambda_0 /2 \over {1 - k \lambda_0/2} }\right)^{k/2} \\ \nonumber
\phi_E &\equiv& {2 \over k \lambda_E} - 1 
\end{eqnarray}
We assume that the bare coupling $\lambda_0$ is sufficiantly weak on the
$1/k$ scale, $\lambda_0 < 2/k$. Then the solution for the running
coupling constant is rewritten as:
\begin{equation}
\label{rgmulti}
\lambda_E =  {\lambda^* \over 
{F^{(-1)}\left[ \left({E \over T_K} \right)^{\Delta} \right] + 1 } }.
\end{equation} 
Here $F^{(-1)}(y)$ is the function inverse to $F(x) = x \exp(x)$. 
The asymptotic form of this solution at $E \ll T_K$ is also useful:
\begin{equation}
\label{lowen}
\lambda_E = \lambda^* - \left( {E \over T_K} \right)^{\Delta}.
\end{equation}

The analysis of the local spin susceptibility is parallel to 
the single-channel case (Section III). It is easy to see that the
uniform part of the local spin susceptibility should vanish in the
multi-channel model as well (see Appendix B). Therefore, 
for the most general magnetic impurity (i.e. with gyromagnetic ratio not 
necessarily 2) we are left with electron and impurity parts of the
oscillating local spin susceptibility $\chi_{2k_F,el}(r,T)$,
$\chi_{2k_F,imp}(r,T)$. The RG equations that these quantities satisfy
were considered in Section II. The only difference with the 
single-channel case is that the $\beta$-function and anomalous dimension 
are different, with $\gamma_{\rm imp}(\lambda)$ now given by:
\begin{equation}
\gamma_{\rm imp}(\lambda) = {k \lambda^2 \over 2}.
\label{blm}
\end{equation}
The non-universal scale factor for the solution of the RG equations
Eq.(\ref{scaling}) then is: 
\begin{equation}
\label{scalm}
\exp\left[-\int_{0}^{\lambda_0}{\gamma_{\rm imp}(\lambda) \over
\beta(\lambda)}d \lambda \right] \simeq {1 \over {1-k\lambda_0/2}}.
\end{equation}
The scaling functions in the large-k limit are determined from the
perturbative analysis (see Appendix A). We find again that  
the scaling equations Eq.(\ref{rg}) are obeyed, and the solutions are 
given by:
\begin{eqnarray}
\label{mc}
\chi_{2k_F,imp}(\lambda_T,x) &=& {1 \over {1-k\lambda_0/2}} 
{\chi^{(1)}_{2k_F}(\lambda_T,x) \over \sinh(2 \pi x)} \\ \nonumber
\chi_{2k_F,tot}(\lambda_T,x) &=& {(3 \pi^3/8) k \lambda_E^2 x \over 
\sinh(2 \pi x)}+{\chi^{(1)}_{2k_F}(\lambda_T,x) \over \sinh(2 \pi x)},
\end{eqnarray}
where 
\begin{equation}
\label{ch1}
\chi^{(1)}_{2k_F}(\lambda_T,x) = (\pi^2/4) 
(1-k\lambda_T/2)^2 {k \lambda_E \over 1 - k \lambda_E/2}             
\end{equation}
and $\lambda_E$ is the coupling at the 
energy scale $E(x)=T/[1-\exp(- 4 \pi x)]e$,
just like in the single-channel case. 
$\lambda_E$, $\lambda_T$ are functions of  
$E/T_K$ or $T/T_K$ given by Eq.(\ref{rgmulti}). Using Eq.(\ref{mc}) together
with Eq.(\ref{rgmulti}) we determine the 
scaling functions $\chi_{2k_F,imp}$ and 
$\chi_{2k_F,el}$ up to the leading order in $1/k$ in the scaling limit 
$\lambda_0 \rightarrow 0$, $r \gg 1/k_F$ at all temperatures:
\begin{eqnarray}
\chi_{2k_F,imp}(T/T_K,x) &=&{\pi^2  \over {2 \sinh(2 \pi x)}} 
\left({F^{(-1)}[(T/T_K)^{\Delta}] \over 
{F^{(-1)}[(T/T_K)^{\Delta}] +1}} \right)^2
{1 \over F^{(-1)}[(E/T_K)^{\Delta}]}\\ \nonumber
\chi_{2k_F,el}(T/T_K,x) &=& {3 \pi^3 x \over
2 k \sinh(2 \pi x)} {1 \over {\left(F^{(-1)}[(E/T_K)^{\Delta}] +1\right)^2}}.
\end{eqnarray} 

It is interesting to note that Eq.(\ref{ch1})
has a factorized form, where the T-dependence is once again that of the spin

\begin{equation}
\label{fctr}
\chi^{(1)}_{2k_F}(\lambda_T,x) = {2 \pi^2 T \chi_{tt}(T) \over
k F^{(-1)}[(E/T_K)^{\Delta}]}.
\end{equation} 

Consider now Eq.(\ref{mc}) in various limits. Obviously,
$\lambda_E = \lambda_r$ for $r \ll \xi_T$, and $\lambda_E=\lambda_T$ for 
$r \gg \xi_T$. At high temperatures $\xi_T \ll \xi_K$, 
and the crossover at $r \sim \xi_K$ does not happen - just like we have seen
in the single-channel case. For $r \gg \xi_T$ the correlation functions
decay exponentially, just as we have seen in the single-channel case.
The most interesting is the low-temperature limit $T \ll T_K$,
$r \ll \xi_T$.  In this limit we find:
\begin{eqnarray}
\label{finsc}
\chi_{2k_F,imp}\left({T \over T_K},{r \over \xi_T} \right) &=&
{\pi \xi_T \over 4 r} {({T/T_K})^{2 \Delta} \over
F^{(-1)}[(\xi_K/[4 \pi r])^{\Delta}]} {1 \over 1 - k \lambda_0/2} \\ \nonumber
\chi_{2k_F,el}\left({T \over T_K},{r \over \xi_T} \right) &=& 
{3 \pi^2 \over 4 k } 
{1 \over {\left(F^{(-1)}[(\xi_K/[4 \pi r])^{\Delta}] +1\right)^2}} + 
(k \lambda_0/2) \chi_{2k_F,imp}\left({T \over T_K},{r \over \xi_T} \right) .
\end{eqnarray} 
The scaling function for the electron piece 
in the limit $\lambda_0 \rightarrow 0$ 
appears in the subleading order in $1/k$. 
For non-zero bare coupling there is also a 
piece in the leading order, 
which is proportional to the impurity scaling 
function in Eq.(\ref{finsc}) and the 
anomalous factor.

As in the single-channel case, the weak 
coupling behavior is \underline{not} recovered
inside the screening cloud. Outside the screening cloud, for $T \ll T_K$  and 
 $\xi_K \ll r \ll \xi_T$, the local spin susceptibility takes the form:
\begin{equation}
\label{ch2}
\chi_{2k_F,tot}\left({T \over T_K},{rT \over v_F}\right) 
= {\pi v_F \over 4 rT  } 
\left({ 4 \pi r T^2 \over  \xi_K T_K^2}\right)^{\Delta} 
+ {3 \pi^2 \over 4 k }.
\end{equation}
The T-divergence is not removed at low temperatures, and Eq.(\ref{ch2}) does 
not have the Fermi-liquid form, as one could expect for a non-Fermi-liquid
low-temperature fixed point.
The distance dependent Knight shift for  
overscreened Kondo fixed point can also be understood at $r > \xi_K$ 
using the generalization
of the Nozi\`eres' Fermi liquid approach developed by one of us and Ludwig \cite{affleck}.

The spin susceptibility is obtained as 
the leading term at the low-temperature fixed point plus corrections in the
leading irrelevant operator.  For the overscreened Kondo fixed
point the leading irrelevant operator 
contribution corresponds to the second term in Eq.(\ref{ch2}). 
It is surprising that the dominant divergent 
term [the first term in Eq.(\ref{ch2})] 
in the limit $T \rightarrow 0$, $r \ll \xi_T$, or in the limit
$k \rightarrow \infty$ appears in the first
order in the leading irrelevant coupling. An interested reader can find the 
details on this technical point in Appendix C.
 
\subsection{Integrated susceptibilities.}

  As we have seen for the single-channel case, the static spin susceptibility
is mostly given by the impurity-impurity correlation function,
$\chi_{ii}(T)$, while other pieces contribute only a small fraction which 
is proportional to the bare coupling constant, or bare coupling constant 
squared. It is easy to see that this is also the case for the 
multi-channel model. Indeed, according to Gan\cite{gan},
\begin{eqnarray}
\chi_{ii}(T) &=& \frac{1}{4T}\,\left(1-k\lambda_0^2\ln\frac{D}{T}\, \right)\\
\nonumber \chi_{ie}(T) &=& - \frac{\lambda_0 k}{8T}\, 
\left(1-k\lambda_0^2\ln\frac{D}{T}\, \right)\\
\nonumber \chi_{ee}(T) &=&  \frac{\lambda_0^2 k^2}{16T}\, 
\left(1-k\lambda_0^2\ln\frac{D}{T}\, \right).
\end{eqnarray}
Thus, from the scaling equations Eq.(\ref{rg}) we obtain:
\begin{equation}
\chi_{ii}=\frac{\chi_{tt}}{(1-k\lambda_0/2)^2}\, \ \ 
\chi_{it}=\frac{\chi_{tt}}{1-k\lambda_0/2},
\label{chitm0}
\end{equation}
where the scaling function for the total spin susceptibility is given by:
\begin{equation}
\chi_{tt} = \frac{1}{4T}\, \left(1-\frac{k \lambda_T}{2}\, \right)^2.
\label{chitm}
\end{equation} 
Using Eq.(\ref{rgmulti}), we can rewrite Eq.(\ref{chitm}) 
in the form:
\begin{equation}
\chi_{tt}(T) = {1 \over 4 T} 
\left({F^{(-1)}[(T/T_K)^{\Delta}] \over {F^{(-1)}[(T/T_K)^{\Delta}] +1}}
\right)^2.
\end{equation}
The electron-impurity and impurity-impurity correlators 
contain smallness associated with the bare coupling:
\begin{equation}
\chi_{ie} \simeq - \frac{k \lambda_0}{2} \chi_{tt}(T) \ \
\chi_{ee} \simeq \frac{k^2 \lambda_0^2}{4} \chi_{tt}(T).
\label{partsm}
\end{equation} 
Thus, the spin susceptibility is given mostly by the impurity-impurity 
spin correlator, and for a system with impurity g-factor $g \neq 2$
there are corrections to the bulk susceptibility proportional to 
the bare coupling. 
For $T<T_K$ the scaling function for the total spin susceptibility 
takes the form:
\begin{equation}
\chi_{tt} \simeq \frac{1}{4 T}\, \left(\frac{T}{T_K}\,\right)^{2\Delta}.
\label{chitlow}
\end{equation} 
As in Section IIIB, this fact is easily understood 
in the bosonic language using canonical transformation.
The bosonized Kondo Hamiltonian for the $k$-channel model has the form:
\begin{eqnarray}
H &=& \int_{-\infty}^{\infty} dr 
\left[ (\partial_r \varphi_L(r) )^2 - {h_e \sqrt{k} \over
\sqrt{2 \pi}} \partial_r \varphi_L(r) \right] + H_0^{para} 
+ \sqrt{2 \pi k} \lambda_0 S^z \partial_r \varphi_L(0) 
\nonumber \\
& & + const \cdot \lambda_0 ( S^+ e^{i \sqrt{8 \pi/k} \varphi_L(0)} O^{para} 
+ h.c.) - h_i S^z.
\end{eqnarray}
Here $\varphi$ is  the canonically normalized total spin boson, i.e. the 
sum of the spin bosons for each channel divided by $\sqrt{k}$.  The 
additional, independent degrees of freedom which couple to the impurity 
correspond to the $SU(2)_k$ Wess-Zumino-Witten model with one free boson 
factored out.  This is the $Z_k$ parafermion 
model\cite{Zamolodchikov}.  For the $k=2$ case it corresponds to an 
extra Ising degree of freedom, or equivalently a Majorana fermion.  
These extra degrees of freedom play no role in
the canonical transformation.

Changing the bosonic field $\varphi_L(r) = \tilde{\varphi}_L(r) + 
{\sqrt{k} h_e r \over \sqrt{8 \pi}}$,
the Hamiltonian takes the form:
\begin{eqnarray}
H &=& \int_{-\infty}^{\infty} dr (\partial_r \tilde{\varphi}_L(r) )^2 
+ H_0^{para} - {k \over 8 \pi} h_e^2 (2 L) + \sqrt{2 \pi k} \lambda_0
S^z \partial_r \tilde{\varphi}_L(0)  
\nonumber \\
& &  + const \cdot \lambda_0 ( S^+ 
e^{i \sqrt{8 \pi/k} \tilde{\varphi}_L(0)} O^{para} 
+ h.c.) - \left(h_i k \lambda_0 {h_e \over 2}\right) S^z .
\end{eqnarray}
Thus, for the Free energy, we have:
\begin{equation}
F(h_i,h_e) = - {Lk \over 4 \pi v_F}  h_e^2  + F(0, h_i- {h_e k \lambda_0 \over 2})
\end{equation}
For the  susceptibilities we then obtain:
\begin{equation}
\chi_{ie} = - {k \lambda_0 \over 2} \chi_{ii},  \ \ \chi_{ee} = 
\left({k \lambda_0 \over 2} \right)^2 \chi_{ii},
\end{equation}
with $\chi_0 = {k \over 2 \pi v_F} L$, the Pauli term. This agrees with the
large-k results.

Let's now return to the issue of screening.
The electron-total piece of the spin susceptibility,
$\chi_{et}=\chi_{ee} + \chi_{ei}$, is given by the integral of the local
spin susceptibility $\chi(r,T)$. As in the single-channel case,
since $\chi(r,T)$ only has the
oscillating piece, this integral is determined by the short-distance
contribution, $r \sim 1/k_F$. The form of $\chi(r,T)$ at $r \sim 1/k_F$ is cutoff-dependent.
However this dependence disappears in the integral, which describes conduction
electron spin polarization. 
In case of a 3D Fermi gas the cutoff procedure is well-defined.
The fact that the net conduction 
electron spin polarization due to impurity comes mainly
from $r \sim 1/k_F$ is indeed 
justified to the orders we worked in perturbation theory.
From Eq.(\ref{mc}), 
with $\lambda_E \simeq \lambda_r \simeq \lambda_0 \lesssim 1/k$
we can write for the local spin susceptibility:
\begin{equation}
\label{3dch}
\chi({\bf r},T) = {k \lambda_0  \chi_{tt}(T) \over  1 - k \lambda_0/2} 
\left({\cos{2 k_F r} \over 
8 \pi r^3} - {\sin{2 k_F r} \over 16 \pi k_F r^4} \right).
\end{equation}      
We have checked this conjecture to the leading order in $1/k$.
Integration of this expression over $r$ 
gives the correct result for $\chi_{et}$,
\begin{equation}
\chi_{et}(T) = {- k \lambda_0/2 \over 1- k \lambda_0/2} \chi_{tt}(T).
\end{equation}
 Obviously, the major contribution to the integral 
\begin{equation}
\label{iii}
\int_0^{\infty} d^3{\bf r} \chi({\bf r}, T) \propto 
\int_0^{\infty} d r {d \over dr} \left({\sin{2 k_F r} \over 2 k_F r}\right)
\end{equation}
comes from $r \sim 1/k_F$.

\subsection{Equal-time correlation function.}       

As we have seen above, 
the zero-frequency spin correlator vanishes as $T^{2 \Delta}$ when
$T \rightarrow 0$. It also obeys a zero-sum rule.
 As in the single-channel case,
the equal-time spin correlator  
$K(r,T)=\left<S_{el}^z(r,0)S_{imp}^z(0)\right>$ has
a nonvanishing sum rule, since $\left<S^z_{el} S^z_{imp}\right> = -1/4$.
The uniform part of the equal-time spin correlator is non-zero.

Consider the equal-time correlators $K_{un}(r,T)$ and $K_{2k_F}(r,T)$ using
the $1/k$ expansion. K(r,T) satisfy scaling equations Eq.(\ref{rg}).
As in the single-channel case, 
for $r > \xi_T$ the spin correlators decay exponentially. 
The behavior is most interesting for $r \ll \xi_T$,
where our expressions are considerably simplified.
Expressing our results in terms of 
effective coupling at scale $r$, $\lambda_r$, we get:
\begin{eqnarray}
\label{meq}
K_{un}(\lambda_r, r/\xi_T) &=& 
- {1 \over 1 - k \lambda_0/2} {\pi v_F k \lambda_r^2 \over 2 r}
(1-k\lambda_r/2)^2 \\ \nonumber 
K_{2k_F}(\lambda_r, r/\xi_T) &=& 
{1 \over 1 - k \lambda_0/2} {\pi v_F \over 8 r} k \lambda_r 
(1 - k \lambda_r/2)
\end{eqnarray} 
We can rewrite these expressions using Eq.(\ref{rgmulti}) in terms of 
$T/T_K$, $r/\xi_T$ variables. 
Suppressing the anomalous factor $1/(1-k\lambda_0/2)$, we obtain:
\begin{eqnarray}
\label{ku}
K_{un}\left({T \over T_K}, {r \over \xi_T} \right)  &=& 
- {2 \pi  T \xi_T 
\over k r} L^2[(4 \pi r/\xi_K)^{\Delta}] \\ \nonumber
K_{2k_F}\left({T \over T_K}, {r \over \xi_T} \right) &=& 
{\pi T \xi_T  \over 4 r} L[(4 \pi r/\xi_K)^{\Delta}] ,
\end{eqnarray}
where $L(x)$ is the function defined by
\begin{equation}
L(x) \equiv {F^{(-1)}(1/x) \over (F^{(-1)}(1/x) + 1)^2}.
\end{equation}
A plot of this function is shown in Fig. \ref{Lx}.
As we have discussed in Section III, the integral of $K(r,T)$ 
should not vanish. It is
given by Eq.(\ref{smr}), as in the single-channel case. 
The integral over long distances $r \sim \xi_K$ 
can be calculated explicitly from Eq.(\ref{meq}) by changing variable 
$r \rightarrow \lambda_r$. Using Eq.(\ref{bedef}),
\begin{equation}
 \int_0^{\infty} {dr K_{un}(\lambda_r,r) \over 2 \pi v_F} =
\int_{\lambda_0}^{\lambda^*} 
{k \lambda_r-2 \over 2 - k \lambda_0} {d \lambda_r \over 2\lambda^*}
 = {k \lambda_0-2 \over 8}.
\end{equation} 
Thus, in this case
the screening length $\sim \xi_K$ is explicitly present.
The dependence 
on the bare coupling constant $\lambda_0$ is surprising, since it should 
not be there according to the sum rule Eq.(\ref{smr}). The missing part
of the sum rule comes from the short distances. To provide the most 
transparent demonstration of this, we write 
the second equation in Eq.(\ref{meq}) 
for a 3D Fermi gas, so that $\cos(2 k_F r)$
is replaced by 
$\cos(2 k_F r) - [\sin(2 k_F r)/2 k_F r]$, as in Eq.(\ref{3dch}). 
The short-distance integral, which is analogous to Eq.(\ref{iii}),
gives precisely the compensating term $-k \lambda_0/8$ needed for the sum rule
Eq.(\ref{smr}) to be obeyed.  

The low-temperature decay of the equal-time correlator at $r \gg \xi_K$
in the overscreened multichannel Kondo model can be obtained using 
conformal field theory approach (see Appendix C). 
Indeed, at the low temperature fixed point we have\cite{affleck,ludwig}:
\begin{equation}
S_{imp} \to {const \bbox{\phi}(0,0) T_K^{-\Delta}},
\end{equation}
where $\bbox{\phi}$ is the $s=1$ primary of dimension $\Delta = {2 \over 2+k}$,
const is a non-universal constant.
We then obtain for $K_{2k_F}$ from conformal invariance:
\begin{equation}
K_{2k_F}(r) \propto 
\left<{(\bbox{\phi}(0,0) \cdot \bbox{\sigma})\over 2 T_K^{\Delta}} 
\psi_L^{\dagger}(0,r) \psi_L(0,-r) \right>
\propto {1 \over T_K^{\Delta} r^{1+\Delta}},
\end{equation}
in agreement with the large-k result Eq.(\ref{ku}).
The same leading order calculation gives zero for $K_{un}$, since
$\left<\bbox{\phi}(0,0) \cdot {\bf J}(0,r)\right> = 0$. The first-order
term in the leading irrelevant operator gives
\begin{equation}
K_{un}(r) \propto {1 \over T_K^{2 \Delta} r^{1+2 \Delta}},
\end{equation}
which also agrees with Eq.(\ref{ku}). It is interesting to note that,
unlike the single-channel Kondo model, the long-distance decay of the
uniform and $2 k_F$ correlators is different.

\section{Conclusion}

Although the techniques employed in this paper, renormalization group
improved perturbation theory and the large $k$ limit are of limited
validity, they have led to one exact result (all orders in perturbation theory)
and suggested a certain conjecture which, if true, lead
to a rather complete picture of the Kondo screening cloud.  We first
summarize the exact result and the conjecture, pointing out a
consistency check between them and then state the resulting conclusions.
\begin{itemize}
\item{The uniform part of the $r$-dependent susceptibility, vanishes to
all orders in perturbation theory.  On the other hand, the equal time
correlation function has a non-zero uniform part, varying on the scale
$\xi_K$ at $T=0$.} \item{The $2k_F$ part of the $r$-dependent susceptibility
has a factorized form at  $v_F/T \gg r$:  \begin{equation}
\chi_{2k_F}(r,T) \to f(r)\chi_{tt}(T),\end{equation}
where $\chi_{tt}(T)$ is the total susceptibility (less the free electron
Pauli part).  At small $r$ this becomes:
\begin{equation}
\chi_{2k_F}(r,T) \to {k\pi v_F\over
2r}\lambda_r\chi_{tt}(T),\label{fact}\end{equation} where $\lambda_r$ is the
effective coupling at scale $r$.  This was verified to third order in
perturbation theory and in the large $k$ limit (including the  $O(1/k)$
correction).} 
\end{itemize}

There is an important consistency check relating this result and
conjecture and the result $\chi_{ie} = - (\lambda_0/2) \chi_{ii}$, 
following from the formula:
\begin{equation} \chi_{et}=\int d^3r \chi (r).\end{equation}
Since $\chi (r)$ is an RG invariant, it has no explicit dependence on the
bare coupling.  If the uniform part had been non-zero, its integral would
have given a contribution to $\chi_{et}$ which would be unsuppressed by
any powers of the bare coupling.  The integral involving $\chi_{2k_F}(r)$
gives $0$ for $r \gg 1/k_F$ due to the $\cos (2k_Fr)$ factor and hence is
determined by the value of $\chi_{2k_F}$ at short distances of
$O(1/k_F)$.  In this limit $\chi_{2k_F}(r)\propto \lambda_r\approx
\lambda_0$ and integrating Eq. (\ref{fact}) gives
$\chi_{et} \simeq - (\lambda_0 k/2) \chi_{tt}$.

Strictly speaking this consistency check requires yet another conjecture:
\begin{equation}
\chi (r) \approx {\chi_{2k_F}\over 4\pi^2r^2v_F}\left[ \cos (2k_Fr)-{\sin
2k_Fr\over 2k_Fr}\right]\approx {k\lambda_r\chi_{tt}(T)\over 8\pi r^2}
\left[ \cos (2k_Fr)-{\sin
2k_Fr\over 2k_Fr}\right],
\end{equation}
for $r \ll \xi_K, v_F/T$.  This last conjecture, involves corrections of
$O(1/k_F)$ which we have not calculated systematically and go beyond the
scope of the one-dimensional model.  We did check the result in lowest
order in $1/k$.

Despite the limitations of our calculational approach, we are thus led to
a fairly complete understanding of the Kondo screening cloud.  The
heuristic  picture of Nozi\`eres and others of the Kondo groundstate is
seen to be correct.  The impurity essentially forms a singlet with an
electron which is in a wave-function spread out over a distance of
$O(\xi_K)$.  This is seen from our calculation of the $T=0$ equal time
correlation function which varies over the scale $\xi_K$.

On the other hand the behaviour of static susceptibilities is
considerably more subtle.  A naive picture that an infinitesimal magnetic
field fully polarizes the impurity but induces a compensating polarization
of the electrons is certainly wrong.  Rather the impurity polarization is
proportional to the weak magnetic field and the {\it integrated}
polarization of the electrons (with the free 
electron value subtracted) is much smaller (proportional to
$\lambda_0$).  The finiteness of the $T=0$ impurity susceptibility results
from its tendency to form a singlet with the electrons.

If we now examine
the $r$ dependence of the electron polarization, we find that it is small
at short distances ($O(\lambda_0)$). However, it exhibits a universal
oscillating form at long distances which is not suppressed by any powers of
$\lambda_0$ but only by a dimensional factors of $1/r^2$.  The fact that it
is purely oscillating ensures that the contribution to the integrated
polarization is negligible.  The envelope of this oscillating
susceptibility, consists of the dimensional factor of $1/r^2$ times
an interesting and universal scaling function of $r/\xi_K$ and $T/T_K$.
This scaling function factorizes into $\chi_{tt}(T) f(r/\xi_K)$ for
$v_F/T \gg r$.

Our work leaves various open questions for further study.  It seems
plausible that our conjecture could be proven to all orders in
perturbation theory, thus putting this work on a more solid foundation.
There are three interesting universal scaling functions which we have
introduced, one for the $2k_F$ susceptibility and two for the uniform and
$2k_F$ equal time correlation functions.  A general calculation of these
functions could perhaps be accomplished by quantum Monte Carlo or
exact integrability methods.
Results on the $T=0$ limit of the susceptibility scaling function were
given in Ref.[\onlinecite{sorensen}].  
An obvious generalization of our calculations
is to general frequency dependent Green's functions.

Most importantly, experimental results on the Kondo screening cloud are
very limited.  The NMR experiments of Boyce and Slichter only probe
extremely short distances, $r\approx 1/k_F$.  Our work shows that these
results are entirely consistent with a large screening cloud.  However,
these experiments do not directly probe the scale $\xi_K$.  NMR is
probably not a feasible technique for doing this since it is difficult to
study distances of more than a few lattice constants.  One possibility
might be neutron scattering, which could in principle measure $\chi
(q,\omega )$ for $q\approx 2k_F$.  An alternative is to study small
samples with dimensions of $O(\xi_K)$\cite{giordano}.

We would like to thank A.~V.~Balatsky,
J.~Gan, F.~Lesage, N.~Prokof'ev, H.~Saleur, D.~J.~Scalapino, E.~S.~S\o
rensen,   P.~C.~E.~Stamp, B.~Stojkovi\'c, C.~Varma and A.~Zawadowski
for useful discussions and comments. 
This research was supported by NSERC of Canada.

\appendix \section{perturbative Results}

The diagram technique for interactions involving spin operators is 
complicated due to their nontrivial commutation relations. 
It is possible to express these operators in terms of  pseudofermion 
operators\cite{abrikosov,popov}:
\begin{equation}
\label{psferm}
{\bf S}_{imp} = {1 \over 2} \sum_{\alpha,\beta = 1,2} 
f^{\dagger \alpha} \bbox{\sigma}_{\alpha}^{\beta} f_{\beta}.
\end{equation}  
The problem in using the fermion substitution Eq.(\ref{psferm})
is that the $\bbox{\sigma}$-matrices have dimensionality 2, while 
the fermion space is four-dimensional. Thus, only the 
states with 
\begin{equation}
N = \sum_{\alpha} f^{\dagger \alpha}  f_{\alpha} = 1
\end{equation}
are physical. This constraint is imposed by choosing appropriate
chemical potential\cite{abrikosov}. 
For example, Popov's technique\cite{popov} 
adds an imaginary chemical potential, $i\pi T/2$, to the
pseudofermions. Then the contribution of the nonphysical states to the
partition function is zero.
The diagram technique then becomes the standard fermion technique with
the one-dimensional conduction electron (left-movers) propagator 
$(i \omega_n + v_F k)^{-1}$, the pseudofermion 
propagator $(i \omega_n - i [\pi T/2])^{-1}$, and
the interaction Hamiltonian
\begin{equation}
H_{int} = v_F \lambda_0 {\bbox{\sigma}^{\beta}_{\alpha} 
\bbox{\sigma}^{\delta}_{\gamma} \over 4}
\psi_L^{\dagger \alpha}(0) f^{\dagger \gamma}
f_{\delta} \psi_{L \beta}(0).
\end{equation}
For our purpose of computation of spatial correlators it is convenient 
to work in the coordinate $r,\tau$- space, where the 
propagator for the left-movers takes the form:
\begin{equation}
G_0(z) = {\pi T \over \sin{[\pi T z]}}, \ \ z= v_F \tau + i x.
\end{equation}

For the lowest-order diagrams it may be more convenient to 
calculate time-ordered impurity spin averages directly.
Such spin operator
Green's function approach was applied successfully, for example, in case
of long-range Heisenberg ferromagnets\cite{vaks}.
Consider
\begin{equation}
\label{average} 
\left<S^{i} S^{j} .... S^{k}\right>,
\end{equation}
where
$i,j,...,k = \{z,+,-\}$. Obviously, this average is zero when
the total number of $S^+$ operators is not equal to the total number
of $S^-$. Consider first averages containing only $S^z$ operators. 
For odd number of spin operators it vanishes.
In our simple $S=1/2$ case $Tr\left([S^z]^{2n}\right) = 1/4^n$.
One can use spin commutation relations and  the relations
$S^+S^-=(1/2)+S^z$, $S^-S^+=(1/2)-S^z$ to calculate
the average Eq.(\ref{average}).  
                
   All diagrams for the spin 
susceptibility $\chi(r,T)$ up to third order are shown
in Fig.\ref{pert}.   The graphs (a)-(d) represent the electron-impurity
part, while the graphs (e)-(i) the electron-electron part. 
We only show the electron Green functions on these diagrams. The
dashed line represents the boundary. 
For the electron-impurity spin correlation
function the external electron spin operator $S_{el}(r)$ takes the propogator 
away from the boundary. 
In case of electron-electron part of the Knight shift there 
are two such operators. 
We have to integrate over the position of one of these operators.

Straightforward calculations lead to the final results stated
in Eqs. (\ref{Pth}),(\ref{Pthimp}) of Section IIIA. 
To calculate the equal-time correlator 
$\left< S_{el}(r,0) S_{imp}(0) \right>$, 
we need to evaluate the graphs (b)-(d) of Fig. \ref{pert} once again. 
The first graph (a) is frequency-independent, 
i.e. it is the same as for the electron-impurity part of
the local spin susceptibility. Both uniform and $2 k_F$  parts 
are now non-zero. The result of this calculation is given by Eq.(\ref{etp})
of Section IIIC. 

For the discussion of static susceptibilities in 
Section IIIB we need to calculate
impurity-impurity part, in addition to space integrals of $\chi_{ie}(r,T)$
and $\chi_{ee}(r,T)$. The second- and third-order graphs for $\chi_{ii}(T)$
 are shown in Fig.\ref{pert2}. The leading order is, of course, $1/4T$.
We find that
\begin{eqnarray}
\label{apchi}
4 T \chi_{ii} &=& 
1 - \lambda_0^2 \left(\ln {D \over T} + A_1 \right) \\ \nonumber
              &-& \lambda_0^3\left( \ln^2 {D \over T}
+ A_2 \ln {D \over T} + const\right) \\ \nonumber
4 T \chi_{ie} &=& - {\lambda_0 \over 2} + 
B_1 \lambda_0^2 + {\lambda_0^3 \over 2} \ln {D \over T}
+ const \lambda_0^3 \\ \nonumber
4 T \chi_{ee} &=& {\lambda_0^2 \over 4} + const \lambda_0^3.
\end{eqnarray}

In general, the constants $A_1$, $A_2$, 
and $B_1$ in Eq.(\ref{apchi}) depend on the cutoff procedure.
However, these three constants are connected, 
$A_2 + 4 B_1 - 2 A_1 = 0$, as follows directly from the
results of Wilson\cite{wilson} on the scaling properties 
of the total spin susceptibility.
Using this connection and Eq.(\ref{solu}), 
the fact that all three scaling functions for the spin
susceptibility are equal up to the terms $\sim \lambda_T^2$ 
is easily demonstrated.

 Consider now the multi-channel case. As we have mentioned in the text,
the graph selection in this case is different, since each vertex $\lambda$
is $\sim 1/k$. To the order $1/k$ we need to
calculate all the graphs in Fig. \ref{pert}, except (c) and (i),
which are of the order $1/k^2$. In addition, we need to calculate
the fourth-order graph shown in Fig. \ref{graph1}. 
The result of this calculation
is given by Eq.(\ref{mc}) in the text.

Calculations of the equal-time correlator are somewhat more involved.
While $K_{2k_F}(r,T)$ in Eq.(\ref{meq}) is also non-zero up to this
order, $K_{un}(r,T)$ vanishes. We need to go to the next order in $1/k$
to find the answer. 
For the terms of the order $1/k^2$, we need to calculate graph
(c) in Fig. \ref{pert}, and additional fourth and fifth order
graphs shown in Fig \ref{pert1}

The bulk susceptibility results are found again by calculating $\chi_{ii}(T)$
and $r$-integrating the Knight shift. 
In the leading order we only need to consider
second-order graph in Fig. \ref{pert2}.

\section{Proof that the uniform part of the local susceptibility vanishes}
As clarified in the text, the local spin susceptibility can be written as
a sum of impurity and electron parts (see Eq.(\ref{ksh})). We will consider
these two parts separately for the purpose of this proof.

  Consider first the impurity part, $\chi_{un,\rm{imp}}(r)$. Using
Eq.(\ref{Kchi}), one can write:
\begin{eqnarray}
\chi_{un, \rm{imp}}(r,T) & = & v_F \left<T\left[\int_0^{\beta} 
d \tau \psi^\dagger_L(r,\tau){\sigma^z\over 2}\psi_L(r,\tau)
S^z_{\rm imp}(0) exp\left\{-\int_0^{\beta} 
d \tau' H_{int}(\tau')\right\}\right]\right> \\ \nonumber
&+& (r \leftrightarrow -r),
\end{eqnarray}
where $H_{int}$ is given by Eq.(\ref{hamil}).
The fact that this contribution vanishes
is very easily seen when we perform the $\tau$ integration. 
Indeed, in every order in perturbation
theory $\chi_{un, \rm{imp}}(r,T)$ can be written as:
\begin{equation}
\chi_{un, \rm{imp}}(r,T) = 
\int_0^{\beta} \int_0^{\beta} d\tau_1 d\tau_2 I(\tau_1,\tau_2,r)
\times F(\tau_1,\tau_2),
\end{equation}
where
\begin{equation}
I(\tau_1,\tau_2,r) = \int_{0}^{\beta} d\tau G(r,\tau-\tau_1)G(-r,\tau_2-\tau),
\end{equation}
or, equivalently,
\begin{equation}
\label{integral}
I=\int_{0}^{\beta} \frac{d\tau (\pi T)^2}{\sin[\pi T (v_F \tau-v_F \tau_1+ir)]
\sin[\pi T (v_F \tau_2 - v_F \tau - ir)]}.
\end{equation}
After the change of integration variable,
$\tau \rightarrow exp(i 2 \pi T v_F \tau)$, one encounters contour integration with
two poles on one side (see Fig.\ref{proof1}), and $I=0$.

Consider now the electron part, $\chi_{un,\rm{el}}(r)$. Here the
cancellation of $\chi_{un,\rm{el}}(r)$ is less trivial since
there are other graphs in addition to those with the integration
Eq.(\ref{integral}) (see Fig.\ref{proof}):
\begin{eqnarray}
G(r',-\tau_1)G(r-r',\tau)G(-r,\tau_2-\tau)\eta(\tau_2-\tau_1) \\ \nonumber
G(r,\tau-\tau_1)G(r'-r,-\tau)G(-r',\tau_2)\eta(\tau_2-\tau_1),
\end{eqnarray}
where $\eta(\tau_2-\tau_1)$ is determined by the full perturbative series.
We now introduce the complex notation,
$z \equiv \pi T (v_F \tau + i r)$, and remember that $G(z) = \pi T/\sin z$.
Then the sum of the graphs in Fig.\ref{proof} gives:
\begin{eqnarray}
\frac{\eta(z_2-z_1)}{\sin(z-z')}\,\left[\frac{1}{\sin(z_2-z)\sin(z'-z_1)}\, -
\frac{1}{\sin(z_2-z')\sin(z-z_1)}\,\right] & = & \nonumber \\
\frac{\eta(z_2-z_1) \sin(z_1-z_2)}{\sin(z-z_1)
\sin(z_2-z)\sin(z'-z_1)\sin(z_2-z')}, & &
\end{eqnarray}
which is graphically presented in Fig.\ref{proof}. Integration over $\tau$
Eq.(\ref{integral}) yields zero in this case as well. Generalization of this
proof to the multiple number of channels is quite trivial. Indeed, the graphs
that cancel have the same channel-dependent factor.
As we have seen above, the crucial step of the proof is that the integral
Eq.(\ref{integral}) is zero. 
Thus, the $q=0$ part of the correlator is absent only
for zero-frequency spin-spin correlators, not for equal time correlators.

Note that the absence of the uniform part in the distance-dependent Knight 
shift becomes trivial in the bosonic language (see Section IIIB). Indeed, since
\begin{equation}
S^z_{el}(v_F \tau+ir) = {1 \over \sqrt{2 \pi}} \partial_r \varphi_L(v_F \tau+ir),
\end{equation}
we find:
\begin{equation}
\chi_{un}(r) \propto \int_0^{\beta} d \tau \left< {1 \over \sqrt{2 \pi}}
\partial_r \tilde \varphi_L(v_F \tau+ir) S^z \right> 
= - {i \over \sqrt{2 \pi}} (
<\tilde \varphi_L(ir+ v_F \beta) S^z> - <\tilde \varphi_L(ir) S^z>) =0,
\end{equation}
because $\tilde \varphi_L(z)$ is periodic in the imaginary time variable. Note 
that we don't need to worry about a potential short-distance singularity  
because the total spin has been replaced by the impurity spin in the 
expression for $\chi_{un}$ using 
the above argument.  A similar argument for $\chi_{un}=0$ was given in 
Ref.[\onlinecite{lesage}].

\section{Low-temperature long distance local susceptibility 
in the multi-channel Kondo model.}

$\chi_{2k_F}(r,T)$ is determined by the infrared stable fixed point for
$r \gg \xi_K$, $T \ll T_K$ and any value of the ratio $rT/v_F$. For $k>1$
(and $S_{imp}=1/2$) this fixed point is of non-Fermi liquid type.
The low-temperature non-Fermi-liquid 
multi-channel Kondo fixed point was analysed 
by Ludwig and one of us\cite{affleck,ludwig,AL} 
using conformal field theory. We refer
the reader to these works and a recent review\cite{affleck:review} for 
details. In the bosonized form spin, charge, and flavour sectors of 
the free fermion Hamiltonian are separate. 
Only the spin sector is interesting in the
Kondo problem, since the impurity spin couples 
to the spin current. The effect of
the strong coupling fixed point\cite{affleck} is such that the low-temperature
Hamiltonian density is written in terms of new spin currents,
\begin{equation}
\label{lothm}
H_s = {1 \over 2 \pi (k+2)} {\bf J}^2(x),
\end{equation}
where
\begin{equation}
{\bf J}(x) = \sum_j \psi_{Lj}^{\dagger}(x) {\bbox{\sigma} \over 2} \psi_{Lj}(x) + 2 \pi {\bf S}
\delta(x).
\end{equation}
The Fourier modes of the spin currents 
for a system with Hamiltonian density Eq.(\ref{lothm})
defined on a large circle of circumference $2l$,
\begin{equation}
{\bf J}_n = {1 \over 2 \pi} \int_{-l}^l dx e^{i n \pi x/l} {\bf J}(x),
\end{equation}
satisfy the usual Kac-Moody commutation relations,
\begin{equation}
\left[ J^a_n, J^b_m \right] = 
i \epsilon^{abc} J^c_{n+m} + {1 \over 2} k n \delta^{ab}
\delta_{n+m,0}.
\end{equation}
Here $\epsilon^{abc}$ is the antisymmetric 
tensor and $k$ is the Kac-Moody level.
To the leading order, the Knight shift is given by:
\begin{equation}
\chi_{2k_F}(r,T) = 
- {v_F \over 2 \pi}  \int_0^{\beta} \int_{- \infty}^{+ \infty} d \tau d y
\left< \psi_L^{\dagger}(0,r) {\sigma^z \over 2} 
\psi_L(0,-r) J^z(\tau,y) \right>.
\end{equation} 
Using OPE 
\begin{eqnarray}
{{\bf J}(\zeta) \bbox{\sigma} \over 2 } \psi_L(z) &=&  
- {3/4 \over \zeta -z} \psi_L(z) + Reg(\zeta-z) \\ 
{{\bf J}(\zeta) \bbox{\sigma} \over 2 } \psi^{\dagger}_L(z) &=& 
{3/4 \over \zeta -z} \psi^{\dagger}_L(z) + Reg(\zeta-z),
\end{eqnarray}      
where $Reg(\zeta-z)$ denotes a function 
which is regular at $\zeta \rightarrow z$,
we rewrite $\chi_{2k_F}(r,T=0)$ as
\begin{equation}
\label{xxxx}
\chi_{2k_F}(r) = 
- {v_F \over 8 \pi} \int_{- \infty}^{+ \infty} \int_{- \infty}^{+ \infty}
d \tau d y \left[- {1 \over \tau + iy-ir}  
+ {1 \over \tau + iy+ir} \right] <\psi_L(0, -r)
\psi_L^{\dagger}(0,r)>.
\end{equation}
The Green's function for two points on the 
opposite sides of the boundary takes the form:
\begin{equation}
\left< \psi^{\dagger}_L(z_1) \psi_L(\bar{z}_2) \right> = 
{S_{(1)} \over z_1 - \bar{z}_2},
\end{equation}
where
\begin{equation}
 S_{(1)} = {\cos[2\pi/(2+k)] \over \cos[\pi/(2+k)]}
\end{equation}
is the S-(scattering) matrix, calculated in Ref.[\onlinecite{AL}]. 
This is a universal
complex number, which depends on the universality 
class of the boundary conditions.
In the one-channel Kondo effect $S_{(1)}=-1$, 
corresponding to  a $\pi/2$ phase shift.
At the overscreened Kondo fixed points $|S_{(1)}|<1$, 
which means  multiparticle
scattering. Substracting free electron 
contribution and performing the integrals,
we find:
\begin{equation}
\label{NFLR}
\chi_{2 k_F}(r) = k{1 - S_{(1)} \over 2}.
\end{equation}
In the limit $k \rightarrow \infty$ this 
gives $\chi_{2k_F}(r) \simeq 3 \pi^2/4 k$,
in agreement with the large-k result of Section IV. 
For $k=1$ it agrees with the Fermi liquid
result Eq.(\ref{FLR}). Note that no anomalous power 
laws occur in the leading order in
irrelevant coupling constants. Only the 
normalization reflects the non-Fermi liquid behavior.
As in the single-channel case,
finite temperature calculations multiply this expression by the factor 
$2 \pi x/\sin(2 \pi x)$, where $x=rT/v_F$.

Consider now corrections to this expression. The leading
irrelevant operator which appears\cite{affleck} in the effective  lagrangian 
at the overscreened Kondo fixed point is ${\bf J}_{-1} \cdot \bbox{\phi}$, where $\bbox{\phi}$
is the $s=1$ $SU(2)$ KM primary field with the dimension $\Delta = 2/(2+k)$.
The dimension of this singlet operator is $1 + \Delta$. We can again 
write this additional piece as
\begin{equation}
H_{int} \sim {1 \over T_K^{\Delta}} ({\bf J}_{-1} \cdot \bbox{\phi}(0)).
\end{equation}
Thus the correction is given by:
\begin{equation}
\label{haha}
\delta \chi_{2k_F}(r,T) = 
- {v_F \over 2 \pi T_K^{\Delta}}  \int_0^{\beta} \int_0^{\beta}
 \int_{- \infty}^{+ \infty} d \tau d \tau_1 d y
\left< \psi_L^{\dagger}(0,r) 
{\sigma^z \over 2} \psi_L(0,-r) J^z(\tau,y) ({\bf J}_{-1} \cdot
\bbox{\phi}(\tau_1,0))\right>.
\end{equation}
To find the  most singular part of this 
expression as $r \rightarrow 0$, we use the
boundary OPE
\begin{equation}
\label{bope}
\psi_L^{\dagger}(0+ir){\bbox{\sigma} \over 2} 
\psi_L(0-ir) \rightarrow {C \bbox{\phi}(0,0) \over r^{1-\Delta}}.
\end{equation}
From conformal invariance, this zero-temperature correlator 
\begin{equation}
\label{corrr}
\left<(\bbox{\phi}(0) \cdot {\bf J}(z_1))({\bf J}_{-1} \cdot 
\bbox{\phi}(z_2))\right>
= {C' \over |z_1|^{2 \Delta} |z_1-z_2|^2 |z_2|^{2 \Delta}}.
\end{equation}
The finite-temperature correlation function 
which appears under the integral in 
Eq.(\ref{haha}) can be obtained using conformal mapping, 
a conformal transformation which maps the finite-temperature 
geometry (half-cylinder)
onto the zero-temperature half-plane.
\begin{equation}
z=\tan{(\pi T w)}.
\end{equation}
Here $w=\tau+ir$ in the finite-temperature geometry. 
A Virasoro primary operator
$A(z)$ of left scaling dimension $\Delta_A$ transforms as
\begin{equation}
A(w) = \left({dw \over dz } \right)^{-\Delta_A} A(z),
\end{equation}
under conformal transformation.
Using $dw(z)/dz=1/\pi T (1+z^2)$, we express 
the finite-temperature  correlators
in terms of the zero-temperature ones.  The net effect is such that 
the factors $1/(z_1-z_2)$ for the half-plane get replaced by
$\pi T/\sin(\pi T [w_1-w_2])$ on the half-cylinder. Doing the integral in 
Eq.(\ref{haha}) and dropping the constants, we obtain:
\begin{equation}
\delta \chi_{2 k_F}(r,T) \propto 
{1 \over r^{1-\Delta} T^{1-2\Delta} T_K^{\Delta} }, 
\end{equation}
in agreement with the large-k result of Section IV,
This term is subdominant, for $r \ll v_F/T$, compared to
the leading term in Eq.(\ref{NFLR}). On the other hand, it becomes larger
than the ``leading'' term if we take $T \rightarrow 0$ with $r \gg \xi_K$
held fixed. Anomalous powers appear from irrelevant operator corrections.


\newpage

\begin{figure}
\centerline{\epsfxsize=4in\epsfbox{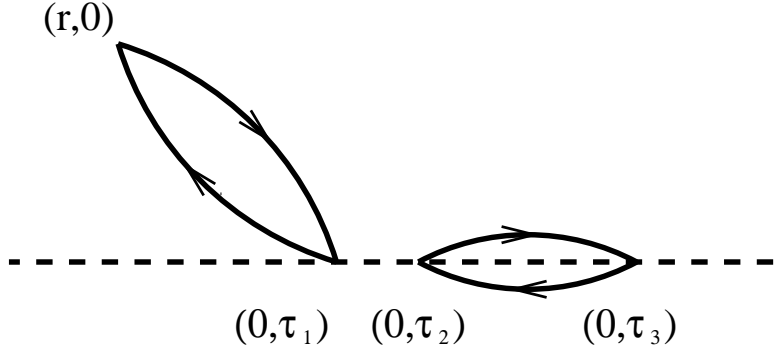}}
\vspace{3mm}
\caption{Singular third-order graph for $\chi(r,T)$.}
\label{graph}
\end{figure}

\begin{figure}
\centerline{\epsfxsize=4in\epsfbox{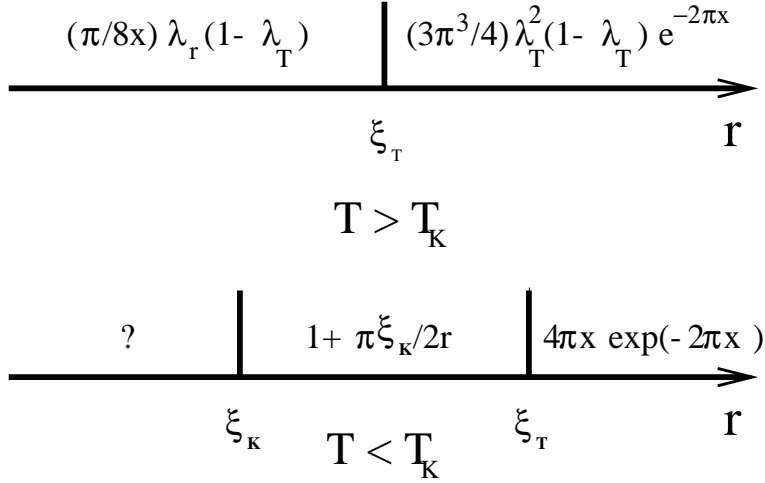}}
\vspace{3mm}
\caption{Scaling regimes for $\chi_{2k_F}(\lambda_T,x=r/\xi_T)$ in
 the single-channel Kondo effect.}
\label{diagr1}
\end{figure}

\begin{figure}
\centerline{\epsfxsize=4in\epsfbox{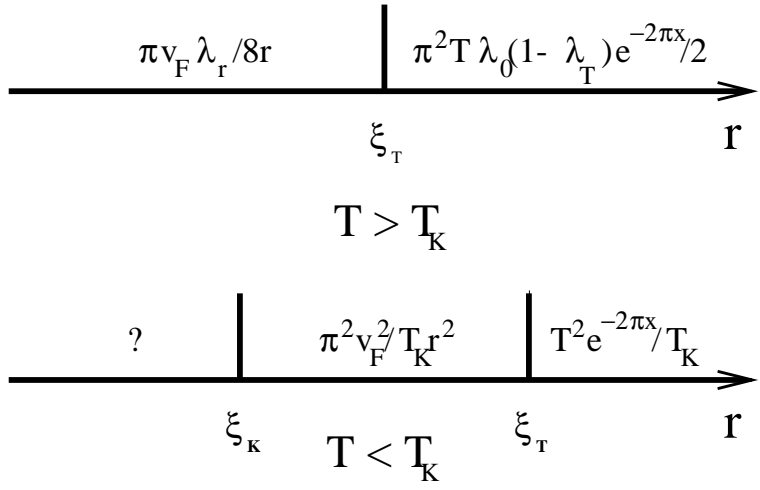}}
\vspace{3mm}
\caption{Scaling regimes for the oscillating part of the equal-time
 spin-spin correlator $K_{2k_F}(\lambda_T,x=r/\xi_T)$ in
 the single-channel Kondo effect.}
\label{diagr2}
\end{figure}

\begin{figure}
\centerline{\epsfxsize=4in\epsfbox{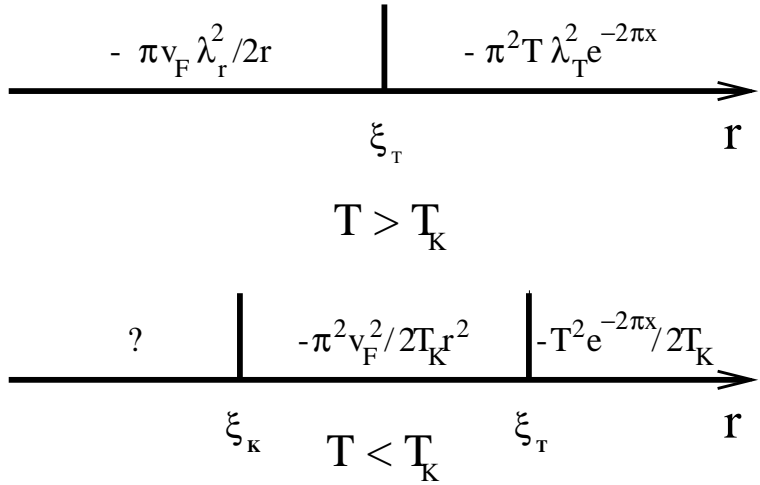}}
\vspace{3mm}
\caption{Scaling regimes for the uniform part of the equal-time
correlator $K_{un}(\lambda_T,x=r/\xi_T)$ in
 the single-channel Kondo effect.}
\label{diagr3}
\end{figure}

\begin{figure}
\centerline{\epsfxsize=4in\epsfbox{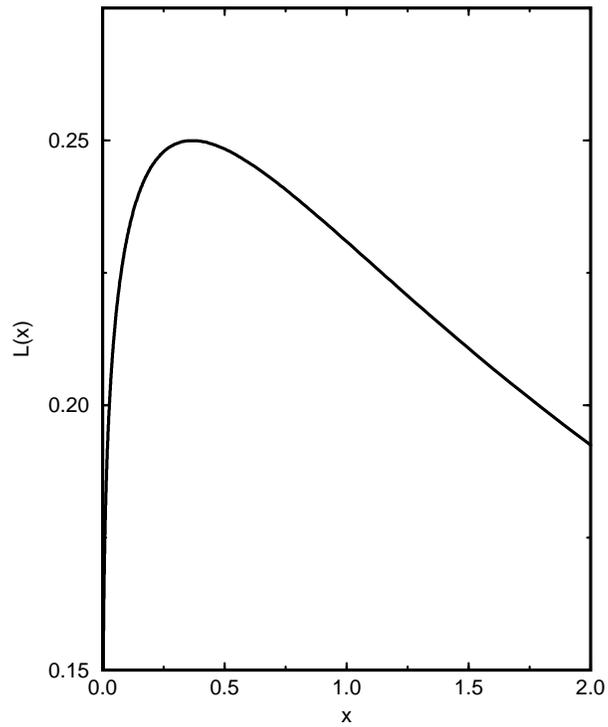}}
\vspace{3mm}
\caption{Scaling function L(x).}
\label{Lx}
\end{figure}

\begin{figure}
\centerline{\epsfxsize=4in\epsfbox{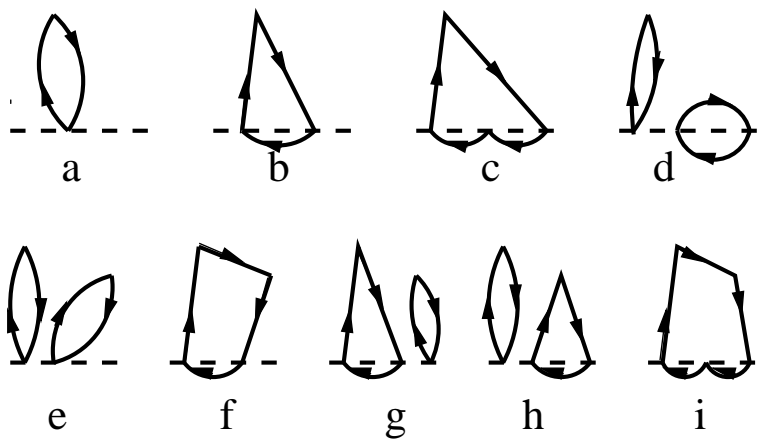}}
\vspace{3mm}
\caption[pert]{Perturbative diagrams for $\chi(r,T)$ up to third order.}
\label{pert}
\end{figure} 

\begin{figure}
\centerline{\epsfxsize=4in\epsfbox{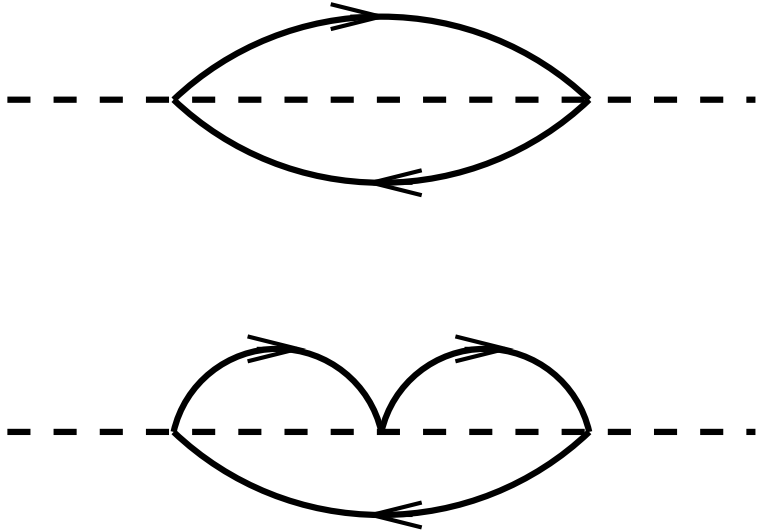}}
\vspace{3mm}
\caption[pert2]{Second and third-order graphs for impurity-impurity part of
the spin susceptibility, $\chi_{ii}$}
\label{pert2}
\end{figure}  

\begin{figure}
\centerline{\epsfxsize=4in\epsfbox{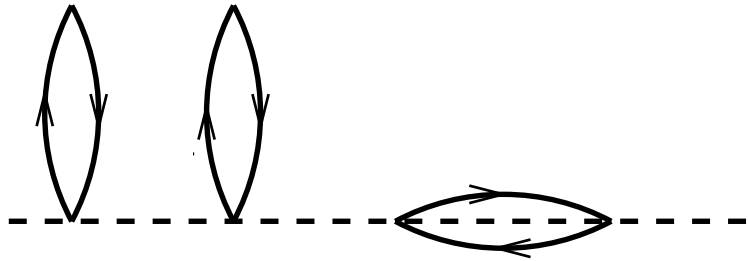}}
\vspace{3mm}
\caption[pert]{Fourth-order graph of the order $1/k$.}
\label{graph1}
\end{figure}  

\begin{figure}
\centerline{\epsfxsize=3in\epsfbox{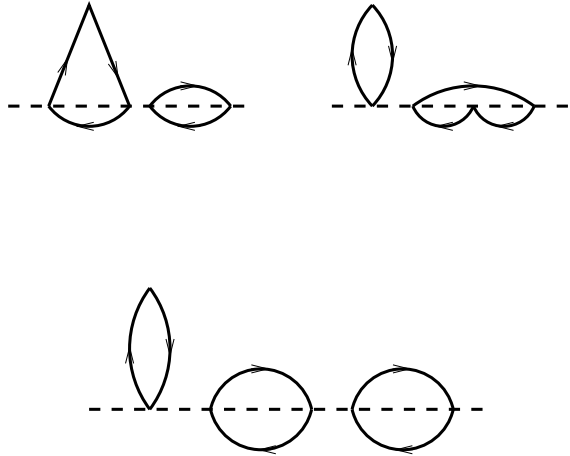}}
\vspace{3mm}
\caption[pert]{Fourth and fifth order graphs for $\chi(r,T)$ that contribute
to the order $1/k^2$ in the $1/k$ expansion.}
\label{pert1}
\end{figure}

\begin{figure}
\centerline{\epsfxsize=4in\epsfbox{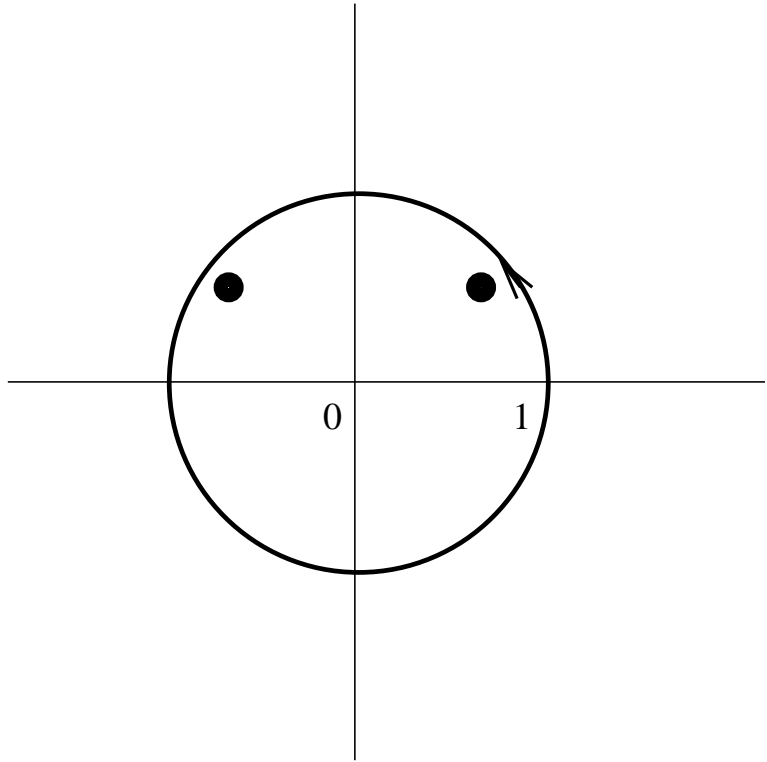}}
\vspace{3mm}
\caption[proof1]{Contour of integration for I.}
\label{proof1}
\end{figure}

\begin{figure}
\centerline{\epsfxsize=4in\epsfbox{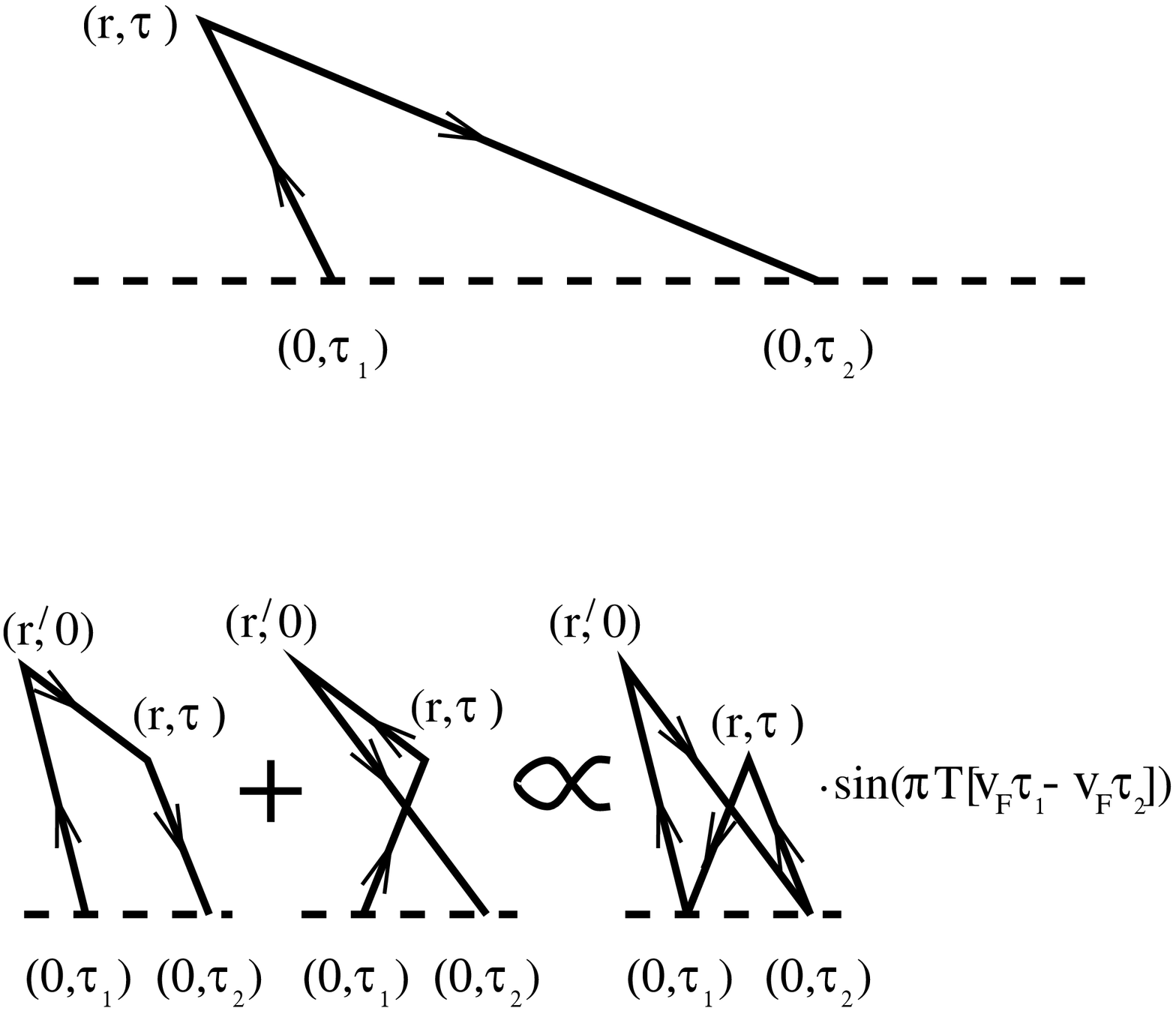}}
\vspace{3mm}
\caption[proof]{Cancellation of the uniform part of the local
spin susceptibility.}
\label{proof}
\end{figure}

\begin{references}
\bibitem{hewson}
A.~C.~Hewson, {\it The Kondo Problem to Heavy Fermions},
Cambridge University Press, Cambridge 1993.
\bibitem{kondo}
J.~Kondo, Solid State Physics {\bf 23}, 183 (1969).
\bibitem{nozieres}
(a) Ph.~Nozi\`eres, J. Low Temp. Phys. {\bf 17}, 31 (1974);
Ph.~Nozi\`eres, J. de Phys. {\bf 39}, 1117 (1978);
{}(b) For a review see Ph.~Nozi\`eres, in {\it Proc. 14th Int. Conf. on Low
Temperature Physics}, eds. M.~Krusius and M.~Vuorio, Vol. 5, (North
Holland, Amsterdam 1975).
\bibitem{gan}J.~Gan, J. Phys.:Cond. Mat. {\bf 6}, 4547 (1994).
\bibitem{sorensen}
E.~S.~S\o rensen and I.~Affleck,  
Phys.  Rev. B {\bf 53}, 9153 (1996); No. cond-mat/9508030.
\bibitem{BA} V.~Barzykin and I.~Affleck, 
Phys. Rev. Lett. {\bf 76}, 4959 (1996).
\bibitem{zawadowski}
O.~\'Ujs\'aghy, A.~Zawadowski, and B.~Gyorffy, Phys. Rev. Lett. {\bf 76},
2378 (1996).
\bibitem{Cox} E.~Kim, M.~S.~Makivik and D.~L.~Cox, Phys. Rev. Lett. {\bf 75},
2015 (1995); E.~Kim and D.~L.~Cox, cond-mat/9706113.
\bibitem{slichter}
J.~P.~Boyce and C.~P.~Slichter, Phys. Rev. Lett. {\bf 32}, 61 (1974);
Phys. Rev. B {\bf 13}, 379 (1976).
\bibitem{Anderson} P.~W.~Anderson, Phys. Rev. {124}, 41 (1961).
\bibitem{schrieffer} P.~W.~Anderson, G.~Yuval, in {\it Magnetism}, v.{\bf 5},
p. 217, eds G.~T.~Rado  and H.~Suhl, Academic Press, London, 1973; 
D.~R.~Hamann, J.~R.~Schrieffer, ibid., p.237.
\bibitem{lowenstein}
J.~H.~Lowenstein, Phys. Rev. B {\bf 29}, 4120 (1984).
\bibitem{lesage}  F.~Lesage, H.~Saleur, Nucl. Phys. {\bf B},
to appear, cond-mat/9611025.
\bibitem{zachar} O.~Zachar, S.~A.~Kivelson, and V.~J.~Emery, 
Phys. Rev. Lett. {\bf 77}, 1345 (1996).
\bibitem{early} M.~S.~Fullenbaum and D.~S.~Falk, 
Phys.Rev. {\bf 157}, 452 (1967); 
A.~P.~Klein, Phys. Rev. {\bf 181}, 579 (1969); 
H.~Keiter, Z. Physik {\bf 223}, 289 (1969).
\bibitem{chen}
K.~Chen, C.~Jayaprakash and, H.~R.~Krishnamurthy,
Phys. Rev. B {\bf 45}, 5368 (1992).
\bibitem{affleck}
I.~Affleck and A.~W.~W.~Ludwig, Nucl. Phys. B {\bf 360},
641 (1991).
\bibitem{ludwig} A.~W.~W.~Ludwig and I.~Affleck, Nucl. Phys. B 
{\bf 428}, 545 (1994).
\bibitem{AL} I.~Affleck and A.~W.~W.~Ludwig, Phys. Rev. B {\bf 48}, 
7297 (1993).
\bibitem{comment} Definitions of $\chi_{2k_F}$ and $\chi_{un}$
differ by a factor of $v_F$ from those used in
Ref.[\protect{\onlinecite{sorensen}}].
\bibitem{migdal}
A.~A.~Abrikosov and A.~A.~Migdal, J. Low. Temp. Phys. {\bf 3},
519 (1970); M.~Fowler and A.~Zawadowski, Sol. State Comm.
{\bf 9}, 471 (1971).
\bibitem{abrikosov} A.~A.~Abrikosov, Physics {\bf 2}, 5 (1965).
\bibitem{wilson}
K.~G.~Wilson, Rev. Mod. Phys. {\bf 47}, 773 (1975).
\bibitem{emery} G.~Toulouse, Phys. Rev. B {\bf 2}, 270 (1970);
M.~Blume, V.~J.~Emery, A.~Luther, Phys. Rev. Lett. {\bf 25}, 450 (1970).
\bibitem{Lesage} F.~Lesage, H.~Saleur and S.~Skorik, 
 Nucl. Phys. B {\bf 474}, 602 (1996).
\bibitem{ishii} H.~Ishii, J. Low. Temp. Phys. {\bf 32}, 457 (1978).
\bibitem{Zamolodchikov} A.B. Zamolodchikov
and V.A. Fateev, Sov. Phys. JETP {\bf 62}, 215 (1985).
\bibitem{giordano}
G.~Chen and N.~Giordano, Phys. Rev. Lett. {\bf 66}, 209 (1991);
M.~A.~Blachly and N.~Giordano, Phys. Rev. B {\bf 46}, 2951 (1992);
J.~F.~DiTusa, K.~Lin, M.~Park, M.~S.~Isaacson, and J.~M.~Parpia,
Phys. Rev. Lett. {\bf 68}, 678 (1992); V.~Chandrasekhar, P.~Santhanam,
N.~A.~Penebre, R.~A.~Webb, H.~Vloeberghs, C.~Van~Haesendonck,
and Y.~Bruynseraede, Phys. Rev. Lett. {\bf 72}, 2053 (1994).
\bibitem{popov} V.~N.~Popov and S.~A.~Fedotov,
Sov. Phys. JETP {\bf 67}, 535 (1988) [ZhETP {\bf 94}, 183 (1988)].
\bibitem{vaks} V.~G.~Vaks, A.~I.~Larkin, and S.~A.~Pikin,
Sov. Phys. JETP {\bf 26}, 188 (1968) [ZhETP {\bf 53}, 281 (1967)].
\bibitem{affleck:review} 
I.~Affleck, Acta Phys. Polon. B {\bf 26}, 1869 (1995).

\end{references}
\end{document}